\pdfoutput=1
\documentclass[10pt,conference]{IEEEtran}
\usepackage{graphicx}
\usepackage{balance}  % for  \balance command ON LAST PAGE  (only there!)
\usepackage{amsmath}
\usepackage{mathtools}
\usepackage{verbatim}
\usepackage{amssymb}
\usepackage{graphicx}
\usepackage{subfigure}
\usepackage{mdwlist}
\usepackage[noend]{algorithmic}
\usepackage{algorithm}
\usepackage{calc}
\usepackage{times}
\usepackage{tabularx}
\usepackage{multirow}
\usepackage[T1]{fontenc}
\newtheorem{defn}{\textbf{Definition}}

\newtheorem{theorem}{\textbf{Theorem}}
\newtheorem{example}{\textbf{Example}}

\begin{document}

\title{Inferring Uncertain Trajectories from Partial Observations}

\author{\IEEEauthorblockN{Prithu Banerjee}
\IEEEauthorblockA{IBM Research\\
  Manyata Tech Park\\
  Bangalore, India\\
prithuba@in.ibm.com}
\and
\IEEEauthorblockN{Sayan Ranu}
\IEEEauthorblockA{Dept. of CSE\\
  IIT Madras\\
  Chennai, India\\
sayan@cse.iitm.ac.in}
\and
\IEEEauthorblockN{Sriram Raghavan}
\IEEEauthorblockA{IBM Research\\
  Manyata Tech Park\\
  Bangalore, India\\
sriramraghavan@in.ibm.com}
}
\maketitle
\begin{abstract}
	The explosion in the availability of GPS-enabled devices has resulted in an abundance of trajectory data. In reality, however, majority of these trajectories are collected at a low sampling rate and only provide partial observations on their actually traversed routes. Consequently, they are mired with uncertainty. In this paper, we develop a technique called \emph{InferTra} to infer \emph{uncertain trajectories} from network-constrained partial observations. Rather than predicting the most likely route, the inferred uncertain trajectory takes the form of an edge-weighted graph and summarizes all probable routes in a holistic manner. For trajectory inference, InferTra employs Gibbs sampling by learning a \emph{Network Mobility Model (NMM)} from a database of historical trajectories. Extensive experiments on real trajectory databases show that the graph-based approach of InferTra is up to $50$\% more accurate, $20$ times faster, and immensely more versatile than state-of-the-art techniques. 
\end{abstract}
\section{Introduction}
\label{sec:introduction}
%Trajectory analytics on road networks is central to a multitude of tasks such as congestion modeling\cite{congestion}, community discovery\cite{lbsn}, and urban resource management\cite{urban}. 
%Generally, a trajectory on a road-network corresponds to a sequence of contiguous nodes, where each node is also tagged with a timestamp denoting the time at which it is traversed. 
The last decade has witnessed an unprecedented growth in the availability of location-tracking technologies, which can be deployed at large scales to collect trajectory data. However, these trajectories are often recorded at a \emph{low sampling rate} wherein the time interval between two consecutive recorded locations is large. As a result, these trajectories only provide partial observations of the actual traversed route and the intermediate portions remain hidden.
%Examples of such low-sampled trajectories include GPS traces from cabs\cite{Wei:2012:CPR:2339530.2339562}, check-in data in social networks\cite{lbsn}, Call Detail Records (CDR)\cite{cdr}, geo-tagged photo albums, etc. As a result, these trajectories only provide partial observations of the actual traversed route and the intermediate portions remain hidden.

Trajectories can be tracked most accurately through gps-enabled devices such as cell-phones or in-car navigations systems. However, a recent work has shown that to reduce power consumption,  majority of the taxis in big cities have a sampling interval exceeding two minutes~\cite{Wei:2012:CPR:2339530.2339562}. The high power consumption of GPS also limits its usage on cell phones over large continuous durations. In the absence of GPS, location of a cell-phone can also be tracked through call detail records (\emph{CDR})~\cite{cdr}, which stores the sequence of base-stations through which a call or data-usage session is routed. However, the problem of low sampling remains since CDRs are generated only when either a call or a data session is in progress. 

High power consumption of GPS is not the sole reason behind low sampling rates. Most social networks today provide ``check-in'' services to announce and share location of a user (e.g., Facebook). Trajectories can be generated from these check-ins by ordering them temporally~\cite{lbsn}. Similar trajectories also arise from geo-tagged photos in photo-sharing sites (e.g., Flickr), credit card transactions and snapshots of vehicles captured through surveillance cameras. Due to the inherent properties of the underlying applications, all of these trajectories are generated at low sampling frequencies. 

\begin{figure}[t]
	\centering
		\includegraphics[width=3.0in]{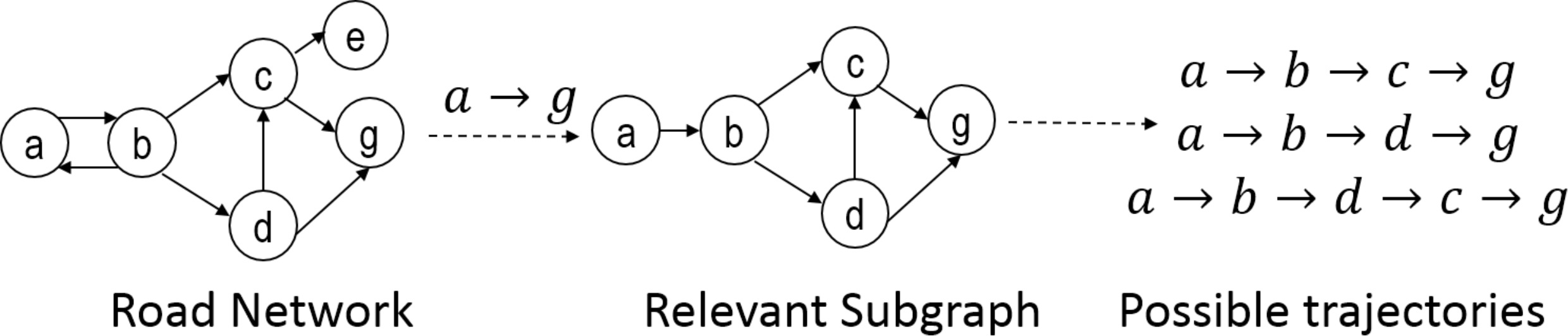}
	\caption{Demonstrates the possible acyclic trajectories arising out of the partial observation $a\rightarrow g$.}
	\label{fig:motivation}
\end{figure}
To understand the uncertainty surrounding low sampling rates, consider the example shown in Fig.~\ref{fig:motivation}. Hereon, we use the term \emph{observations} to indicate a low sampled trajectory, and the term \emph{trajectory} is used to denote the complete sequence of nodes that is actually traversed. The first directed graph shown in Fig.~\ref{fig:motivation} depicts a road network where each node represents a region and an edge corresponds to the road segment connecting two regions. Now, consider the partial observation $a\rightarrow g$. As it can be seen, there is no direct edge from $a$ to $g$. Assuming that trajectories are acyclic, the relevant subgraph in Fig.~\ref{fig:motivation} shows the region within which the mobility is constrained and any of the three paths connecting $a$ and $g$ is a possible trajectory that was actually traversed. 

Managing the uncertainty highlighted in Fig.~\ref{fig:motivation} is critical towards designing accurate higher-order systems that are driven by trajectory data. 
For example, trajectories from geo-tagged photos can be inferred for trip-mining~\cite{Zheng:2009:MIL:1526709.1526816} and used for recommending tourist itineraries. Beyond inferring the most likely trajectory, it is also essential to infer regions which have a high likelihood of being traversed. 
For example, investigative agencies are often interested in retrospective analysis of movements of suspected criminals based on their spatio-temporal footprints generated from CDR data, credit card transactions, surveillance camera snapshots, etc.~\cite{criminal}. Consider a bomb blast that occurred at 9 PM. Investigative agencies are interested in identifying suspected terrorists that were present in the vicinity of the blast site around 9 PM with a high likelihood. For suspects who are mobile, this query cannot be answered using existing mechanisms. Note that the most likely region may not necessarily be part of the most likely trajectory.
%In these investigations, instead of patrolling the most likely trajectory, their goal is to identify all regions that have high likelihoods of being visited and proportionally distribute intelligence gathering resources. 
In essence, we want to capture the entire uncertainty surrounding partial observations, which would allow us to answer the following interesting questions.\\
%The above examples surface the various questions that arise while managing the uncertainty surrounding partial observations. More specifically, given a set of partial observations, we are interested in the following questions.\\
$\bullet$ Which is the most likely trajectory? \\
$\bullet$ What are the top-$k$ most likely road segments? \\
$\bullet$ What are the top-$k$ most likely locations at time $t$?

\begin{figure}[t]
	\centering
		\includegraphics[width=2.49in]{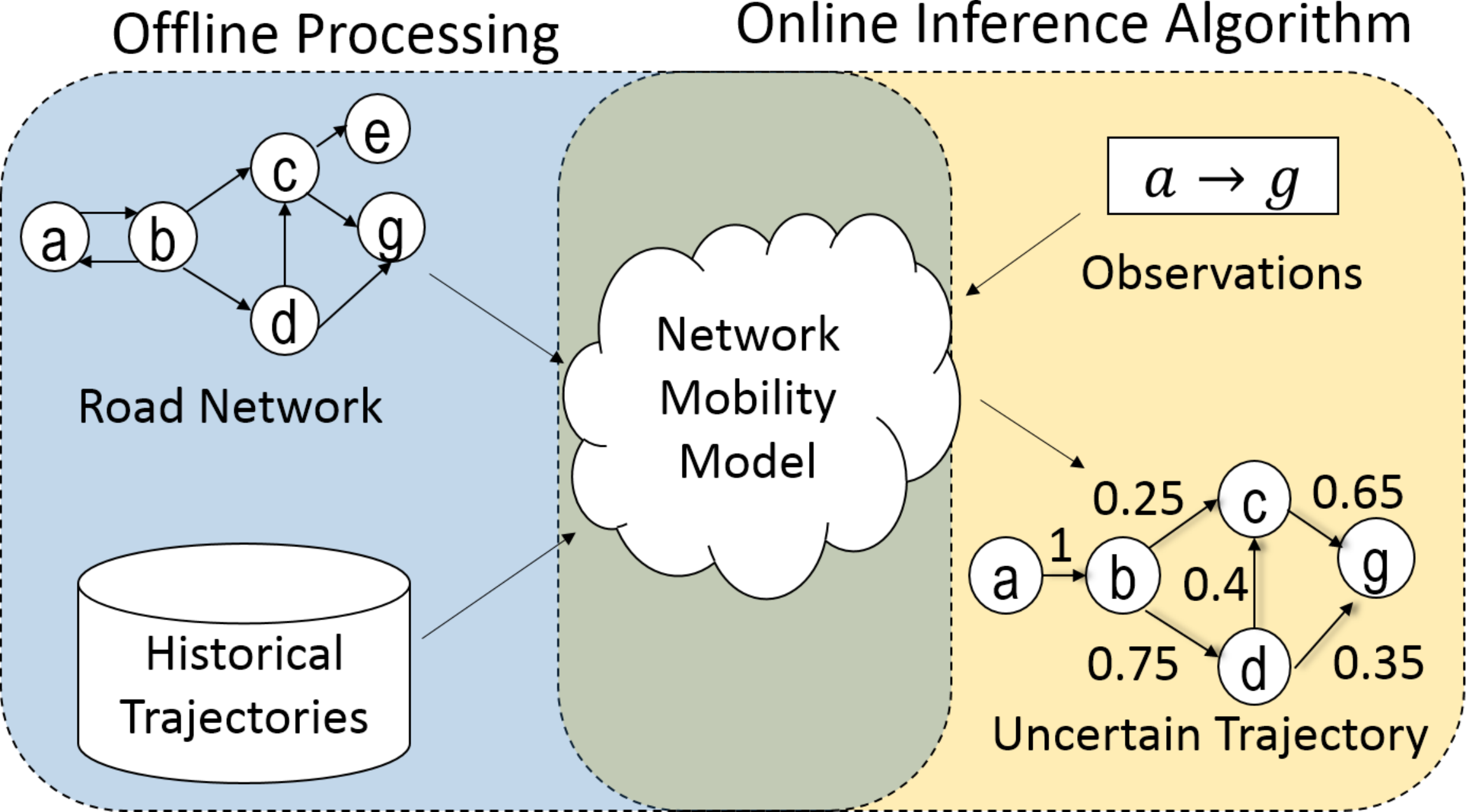}
		\caption{Pipeline of the \emph{InferTra} algorithm.}
		%Additionally, to allow matching uncertain trajectories, an adaptation of the dynamic time warping distance\cite{DTW}, \emph{gDTW}, is proposed.}
	\label{fig:outline}
\end{figure}

In this paper, we design a technique called \emph{InferTra (INFERring TRAjectories)} to answer all of the above questions in a principled manner. Fig.~\ref{fig:outline} outlines the pipeline of InferTra. Against the backdrop of a road network and a database of historical trajectories, InferTra learns a Network Mobility Model (\emph{NMM}), which is then used to predict an ``uncertain'' trajectory. In contrast to existing techniques~\cite{hris}, an uncertain trajectory is an edge-weighted graph. An edge weight denotes the probability of the corresponding road segment being traversed. Overall, the graph summarizes the uncertainty around each possible trajectory arising from the partial observations.
%InferTra ingests two sources of information: the underlying road network, and a database of historical trajectories. From these input data, InferTra learns a Network Mobility Model (\emph{NMM}). Now, given a query set of partial observations, NMM analyzes the observations against the backdrop of the learned NMM, and outputs an edge-weighted graph as an ``uncertain'' trajectory. An  
Using a graph to model the uncertainties is a significant deviation from the current state of the art~\cite{hris}. While one could operate in the world of maximum likelihoods and predict the most likely trajectory, as the uncertainty grows, the information content in maximum likelihood estimations deteriorates. InferTra recognizes this issue and focuses on a more holistic analysis of the uncertainty that imparts both higher accuracy in predictions and the ability to answer a wider range of queries. 
To summarize, the paper makes the following contributions:\\*
 $\bullet$ We develop an algorithm called \emph{InferTra} that infers an ``uncertain'' trajectory from a set of partial observations. Our graph-based approach provides a more holistic representation of uncertainty and thus enabling us to answer deeper questions than state-of-the-art techniques.\\
 $\bullet$ To accomplish inference, we compute a generative model, called Network Mobility Model (NMM), by learning the mobility patterns in a road network from a database of historical trajectories. NMM not only captures the spatial patterns, but it is also the first technique to leverage the temporal signals. \\
 $\bullet$ Empirical evaluations establish the graph-based approach in InferTra as immensely more versatile in answering a wider range of queries. In addition, InferTra is up to 50\%  more accurate and $20$ times faster than HRIS~\cite{hris}.
%As shown later in Sec.~\ref{sec:experiments}, gDTW achieves a higher accuracy than matching only the most likely trajectories.
\section{Related Work}
\label{sec:relatedwork}
HRIS~\cite{hris} is the first and the only work to infer network-constrained trajectories from partial observations. In this section, we outline how InferTra is different. 
%In addition, HRIS showed that map-matching techniques \cite{map1,map2,map3}, which snaps GPS-traces onto an underlying road network, fail to provide accurate predictions at sampling intervals beyond $5$ minutes. Consequently, a pertinent question arises: \textit{What's different in InferTra?} We summarize the differences below.

%In HRIS, the learning happens online where relevant historical trajectories are identified and then analyzed to predict the answer set. On the other hand, InferTra learns the NMM offline and is thus, more scalable. 
$\bullet$ \textbf{1. Inference Goals:} Given a set of input observations, HRIS predicts the most likely trajectory. Thus, it operates in the maximum likelihood world, where the likelihood of a trajectory is modeled using a ``popularity'' score. While the most likely trajectory is a good predictor under low uncertainties, as the uncertainty grows, the information content in the prediction deteriorates. For example, in the uncertain trajectory in Fig.~\ref{fig:outline}, none of the possible trajectories have a high likelihood; rather, they are distributed across the entire relevant subgraph. It is therefore critical to understand how the possibilities that are not captured in the most likely trajectory are spread across the road network. InferTra recognizes this issue. Accordingly, the goal is to infer a single uncertain trajectory that summarizes all of the possibilities in a coherent manner. The uncertain trajectory not only allows us to identify the most likely trajectory, but also  facilitates answering deeper questions such as the most likely location at time $t$, which may not necessarily be part of the most likely trajectory, and identification of all regions with a likelihood above a user-provided threshold. 
Due to this versatility of uncertain trajectories, a decent body of work already exists on querying uncertain trajectories~\cite{uncertain1, uncertain2}, which  is complementary to our problem. Instead of inferring the uncertain trajectory, they assume a database of uncertain trajectories on which further processing is performed.  

%Certainly, one could get a better sense of the uncertainty by predicting the $k$ most likely trajectories in HRIS. However, selecting the optimal $k$ is not straightforward. Second, even if $k$ is known, each predicted trajectory is treated as an independent entity with its own ``popularity'' score. Thus, enough information is not exposed to combine the $k$ trajectories into a graph where scores can be assigned to each road segment or node and answer the advanced queries highlighted above. 
%In contrast, an uncertain trajectory provides a holistic summary, which not only translates to higher prediction accuracies, but also adds versatility in answering both trajectory and node level queries.

$\bullet$ \textbf{2. Capturing Historical Patterns:} HRIS proposes two techniques to connect a pair of consecutive observations. In the first technique, historical trajectories are used to extract a sub-network of the road network within which the mobility is assumed to be bounded. Next, to connect the partial observations, the shortest paths in the sub-network are computed. In the second technique, starting from the first observation, greedy choices are made to iteratively hop to a neighboring node and reach the destination observation. In InferTra, no assumptions, such as preference toward shortest paths, are made based on the network properties. Rather, spatio-temporal patterns displayed by the historical trajectories are learned and utilized in prediction. If indeed shortest paths are favored in certain regions, then this property is automatically learned from the historical trajectories itself.

$\bullet$ \textbf{3. Temporal Signals:} InferTra not only learns the spatial signals embedded in the historical trajectories, but also unearths the temporal signals, which are not utilized in HRIS. 
%As shown later in Sec.~\ref{sec:experiments}, capturing temporal patterns in mobility behaviors is essential for boosting inference performance.

More recently, a technique~\cite{Wei:2012:CPR:2339530.2339562} was designed to study the trajectory inference problem in a setting where trajectories are not constrained by a network. Due to the focus on network-free trajectories,~\cite{Wei:2012:CPR:2339530.2339562} is not applicable to our problem.
\section{Problem Formulation}
\label{sec:formulation}
First, we define the concepts central to our paper.
\begin{defn}
	\label{def:roadnetwork}
	\textsc{Road Network.}  
	\textit{A road network is a directed graph $G(V,E)$. $V$ is the set of nodes representing intersections and terminal points, and $E$ is the set of edges $e=(v_i,v_j)$, connecting $v_i,v_j \in V$, depicting road segments. The position of a node is characterized by its latitude and longitude.}
\end{defn}

We use the notation $e.p_1$ and $e.p_2$ to denote edge $e$'s two endpoints, where $e$ is directed from $p_1$ to $p_2$ where $p_1,\:p_2\in V$.
Generally, a trajectory $T=\{s_1,\cdots,s_n\}$ is a temporally ordered sequence of spatio-temporal points. A spatio-temporal point $s=(v,t)$ is a tuple containing a spatial location $v$ and a timestamp $t$ encoding the time of the day at which $v$ is traversed. We use the notation $T.s_i$ to denote the $i_{th}$ spatio-temporal point in $T$, and $s.v$ and $s.t$ to denote the location and timestamp in $s$ respectively. Indeed, $T.s_i.t<T.s_{i+1}.t$. In this work, we only consider \emph{network-constrained} trajectories.
\begin{defn}
	\textsc{Network-constrained trajectory. }
	\textit{ $T$ is constrained in a road network $G(V,E)$, if $\forall T.s_i,\; T.s_i.v\in V$, and $\forall T.s_i,\; T.s_{i+1}$, $(T.s_i.v,T.s_{i+1}.v)\in E$. }
\end{defn}
\begin{figure}[t]
	\centering
		\includegraphics[width=1.8in]{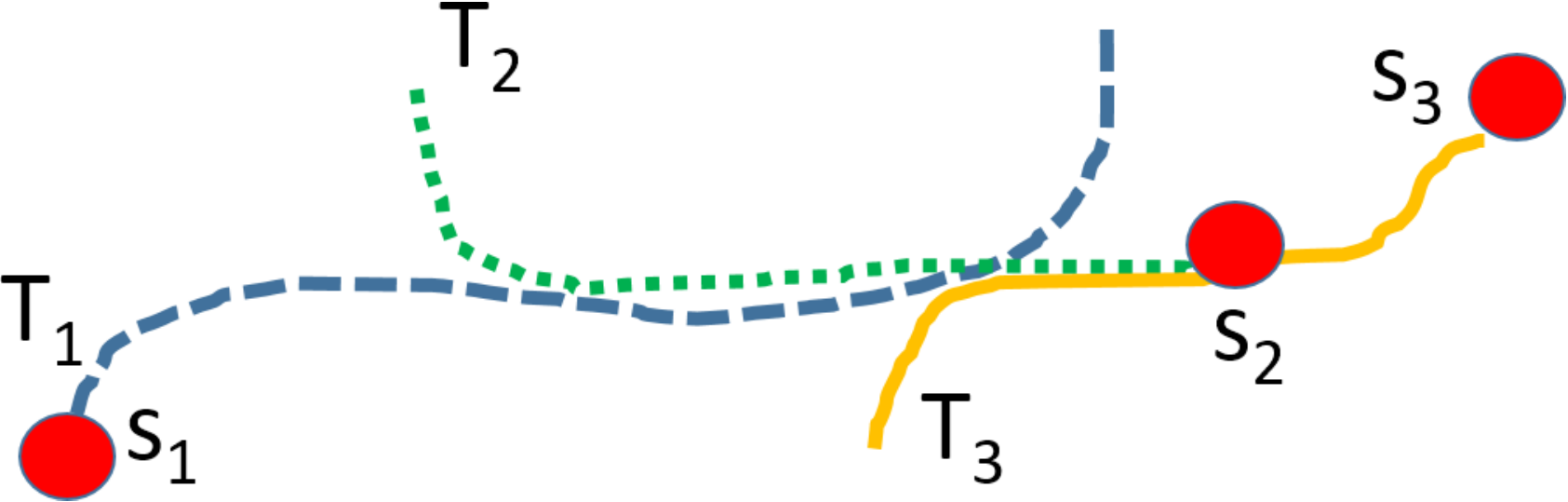}
		\caption{A hypothetical inference scenario.}
	\label{fig:naive}
\end{figure}

In simple words, a network-constrained trajectory is a connected path in the road network. We assume that each edge (or road segment) in a trajectory takes it progressively closer to its destination, and thus a trajectory is acyclic.
%Semantically, a trajectory corresponds to a single directed journey. For example, traveling from home to office in the morning, and then returning to home in the evening are considered two different trajectories. Techniques to partition raw trajectories based on such semantics have been developed\cite{Zheng:2009:MIL:1526709.1526816}. 
%Empirical evaluations in Sec.~\ref{sec:experiments} demonstrate that this assumption generally holds true and does not impact inference performance. 
%Although the movement of vehicles is restricted within a network, trajectories generated from raw GPS traces cannot be assumed to network constrained. This non-conformity of GPS traces arises from various factors such as the sampled point being in the middle of a road segment rather than a node, minor inaccuracies in the location tracking, etc. A class of algorithms\cite{map1,map2,map3}, known as \emph{map-matching}, analyzes this problem and can snap GPS traces to the underlying road network. In our work, we assume that GPS traces have already been map-matched to produce network-constrained trajectories or observations.
As illustrated in Fig.~\ref{fig:motivation}, a set of \emph{partial observations}, is a sequence of network regions where a trajectory $T$, which is unknown, has been spotted. 
\begin{defn}
  \textsc{Partial Observations. }\textit{A partial observation $O=\{s_1,\cdots,s_n\}$, is a temporally ordered sequence of spatio-temporal points where $\forall O.s_i$, $O.s_i.v\in V$.}
\end{defn}
%Certainly $O\subseteq T$. However, the spatio-temporal points in $O$ may not be connected in the road network. 

Given a set of partial observations $O$, our goal is to infer an uncertain trajectory that captures all possible trajectories arising from $O$, quantify the uncertainty associated with each possible trajectory, and capture how this uncertainty is distributed across the road network.   
%Note, $e$ may not necessarily be part of $O$ even if $p(e|O)>0$. One such example is the edge $(b, c)$ in Fig.~\ref{fig:motivation}. The inference problem is now defined as the following.
\begin{defn}
	\textsc{Uncertain Trajectory Inference. }
	\textit{Given a road network $G(V,E)$, a database of historical trajectories $\mathbb{H}$, and a set of partial observations $O$, construct an edge-weighted graph $U(V',E')\subseteq G(V,E)$, such that $E'=\{ p(e|O)>0\;|\;e\in E\}$, where $p(e|O)$ denotes the probability of $e$ being traversed given $O$. $V'$  is defined analogously from the endpoints of edges in $E'$. }
\end{defn}

We assume that the historical trajectories are clean and entirely observed.  

To summarize our formulation, $U$ contains all edges that have a non-zero traversal likelihood, and consequently, encapsulates all possible trajectories. The edge weights capture how the uncertainty is distributed across the road network. Mathematically, $U$ is modeled as a multivariate distribution of its constituent edges, i.e., $p(U|O)=p(e_1,\cdots,e_m|O)$, where $\{e_1,\cdots,e_m\}\in U$.  The edge weights are therefore the marginal distributions, which we need to learn from the observations in conjunction with the evidence provided by the historical data. More precisely, for any given edge $e\in U$,
\begin{equation}
  \label{eq:marginal}
  P(e|O)=\sum_{\forall T\in \mathbb{T}}P(T|O)
\end{equation}
where $\mathbb{T}$ is set of all trajectories (or paths) connecting the sequence of observed nodes in $O$ and also containing, edge $e$. So, if we can compute $p(T|O)$ for all trajectories in $U$, then we can identify all edges $e$ with $p(e|O)>0$, and therefore, compute the uncertain trajectory corresponding to $O$. 
%Since any possible trajectory is a connected path, it is guaranteed that $U$ is connected as well. 
\section{Challenges}
\label{sec:naiveapproach}
Let $O=\{s_1,s_2,s_3\}$ be the observation set and $T=\{e_1,\cdots,e_n\}$ be one path connecting all nodes in $O$. In its simplest form, $p(T|O)$ can be estimated directly from the historical trajectories. More specifically, 
\begin{equation}
  \label{eq:naive}
  p(T|O)=\frac{\Big\|\{T\subseteq T'|T'\in \mathbb{H}\}\Big\|}{\Big\|\{O\subseteq T'|T'\in \mathbb{H}\}\Big\|}
\end{equation}
%we first extract the trajectories $\mathbb{H}_O=\{T\in\mathbb{H}|\:O\subseteq T, T.s_1=O.s_1, T.s_{|T|}=O.s_{|O|}\}$ that start from $O.s_1$, end at $O.s_{|O|}$, and cover all of the intermediate observations in the same order. With a slight abuse of notation, we use $\subseteq$ to denote $O$ as a sub-sequence of $T$. For example, $\{a,c\}\subseteq \{a,b,c\}$. Indeed, one could also include sub-trajectories that satisfy the outlined condition in $\mathbb{H}_O$. For the sake of simplicity we disregard that case since the underlying challenges remain the same. Now, based on $\mathbb{H}_O$,
 
The formulation in Eq.~\ref{eq:naive} works perfectly when the historical database is infinitely large, which, however, is not a realistic assumption. 
%More specifically, it is not possible to represent all possible combinations of input observations in $\mathbb{H}$. Thus, $\{O\subseteq T'|T'\in \mathbb{H}\}\Big\|$ can often be too small, or even null, to estimate the conditional joint distribution $p(T|O)$ directly from $\mathbb{H}$. Consequently, Eq. \ref{eq:naive} can be too restrictive. 
%$ly large database is obtained, the online computation of identifying $\mathbb{H}_O$ will not be scalable. 
Consider the scenario in Fig.~\ref{fig:naive} to understand the limitations of Eq.~\ref{eq:naive} better. For $O=\{s_1,s_2,s_3\}$, no trajectory exists that covers all three regions, and thus, Eq.~\ref{eq:naive} cannot be employed for inference. However, it is easy to see that there are local signals embedded in the historical trajectories that can be consolidated to infer a possible route. More specifically, by stitching together overlapping sub-trajectories from $T_1$, $T_2$, and $T_3$ the likelihood of a path connecting $s_1$, $s_2$ and $s_3$ can be computed. Constructing such arbitrary trajectories from known ones is in fact how a human would draw inference when asked to connect a set of observations that have not been traveled in a single journey. Consider a car that has just crossed $s_2$ in Fig.~\ref{fig:naive} and still needs to find its way to $s_3$. To select the next edge, the driver of the car would ask the following question: \textit{Given my recent past, and based on historical evidence, which road should I take next to maximize my chances of reaching $s_3$?} Thus, the fact that the car started from $s_1$ and there is no path in $\mathbb{H}$ connecting $s_1$ to $s_3$ has no impact on the decision choice between $s_2$ and $s_3$. What matters are the recent past and the current target node to reach, which is $s_3$. 
%Based on the above analysis, a pertinent question arises: \textit{are the historical trajectories useless?} To answer this question, consider the scenario in Fig. \ref{fig:naive}. 

%In other words, we can approximate the globally optimal route by stitching together overlapping locally optimal routes, each of which connects a subset of the observations. 
This natural human tendency of taking locally optimal decisions to construct the globally optimal route can be mathematically expressed as the following. Let trajectory $T$ be the sequence of edges $\{e_1,\cdots,e_n\}$ traveled till now. Then, 
\begin{alignat}{3}
  \label{eq:conditional}
  \nonumber
  p(e_n|e_1,\cdots,e_{n-1},O)&=\\
  \nonumber
  &\hspace{-1.0in}p(e_n | e_{n-m},\cdots,e_{n-1},s_i)\\
  &\hspace{-1.0in}\approx  \frac{\Big\|\{(e_{n-m},\cdots,e_n)\subseteq T | T\in \mathbb{H},\: s_i\in T\}\Big\|}{\Big\|\{(e_{n-m},\cdots,e_{n-1})\subseteq T | T\in \mathbb{H},\: s_i\in T\}\Big\|}
\end{alignat}
\begin{comment}
\begin{alignat}{2}
  \label{eq:conditional}
  \nonumber
  p(e_i|e_1,\cdots,e_{i-1},e_{i+1},\cdots,e_n)&=\\
  \nonumber
  &\hspace{-1.8in}p(e_i | e_{i-m},\cdots,e_{i-1},e_{i+1},\cdots,e_{i+m})\\
  &\hspace{-1.8in}\approx  \frac{\Big\|\{(e_{i-m},\cdots,e_{i+m})\subseteq T | T\in \mathbb{H}\}\Big\|}{\Big\|\{(e_{i-m},\cdots,e_{i-1},e_{i+1},\cdots,e_{i+m})\subseteq T | T\in \mathbb{H}\}\Big\|}
\end{alignat}
\end{comment}
where $m$ quantifies ``recency'' and $s_i\in O$ is an observed node, such that it is not present in $T$, but all its preceding observed nodes, have already been traversed by $T$, i.e., $\forall j,\:s_j\in O,\: 1\leq j<i$; $s_j\in T$.  Thus, Eq.~\ref{eq:conditional} is simply the proportion of historical trajectories that share the same recent history as of $e_n$ and have traveled through $s_i$. 
Now, notice that  Eq.~\ref{eq:conditional} allows us to compute the conditional distribution in an $n$-dimensional space if the $m$-dimensional joint distributions of the numerator and the denominator in Eq.~\ref{eq:conditional} are known. Since $m\ll n$, estimating these $m$-dimensional joint distributions directly from a finite $\mathbb{H}$ is likely to be more accurate. 

To formalize this intuition, let us define the notion of a \emph{density} $\Delta$, which is the ratio of representative samples to the volume of the space. If the density is above a certain threshold $\theta$, then we can assume that the joint distribution sampled directly from the representative samples is accurate. Thus, the density for an $n$-dimensional joint distribution of a trajectory $T$ computed directly from $\mathbb{H}$ is 
 $ \Delta_n=\frac{\|\mathbb{H}\|}{2^n}$. The volume is $2^n$ since there are $n$ random variables and each variable can take two values: traversed or not traversed. So, to satisfy a given threshold $\theta$, the number of representative samples needs to grow exponentially with the dimension of the space. Since $m\ll n$, satisfying this accuracy criteria is significantly easier in an $m$-dimensional space. 
 %conditional can be computed. As $m$ increases, estimating the joint distributions get inaccurate due to exponential increase in the volume of the sample space.
%When $m=n$, it is equivalent to estimating the joint distribution $p(T|O)$ from $\mathbb{H}$ directly, which, as we discussed, is often infeasible.

%While humans can inherently accomplish such tasks, algorithmically, it is not easy. First, the number of trajectories that need to be combined can be arbitrarily large. For example, while two is enough for observation set $\{s_1,s_2\}$, for observation set $\{s_1,s_2,s_3\}$, three trajectories need to be combined. Due to this difficulty, HRIS\cite{hris} limits itself to combining at most two trajectories to extract a candidate subgraph within which further computations are performed. 
The above analysis suggests that, while we may not have the information to compute joint distribution $p(T|O)$ directly, the historical database may be enough to compute conditionals $p(e|T,O)$. Thus, well defined statistical techniques, such as \emph{Gibbs Sampling}~\cite{gibbs}, can be used to approximate the joint distribution by sampling from the conditionals.
\section{Gibbs Sampling for Trajectory Inference}
\label{sec:gibbs}
Gibbs sampling (\emph{GS}) is a generalized probabilistic inference algorithm that is used to generate a sequence of samples from a joint distribution of two or more random variables. The first requirement for GS is some observable data. Let us denote this observed data as $Y$. Next, GS requires a vector of random variables that are unknown to start with. Let us denote this $n$-dimensional vector as $\boldsymbol{\phi}=(\phi_1,\cdots,\phi_n)$. The goal of GS is to learn $\boldsymbol{\phi}$ to model the observable data. Towards that goal, GS follows the following iterative procedure.
\begin{enumerate*}
  \item Initialize each $\phi_i\in\boldsymbol{\phi}$ to some arbitrary value.
  \item for $\tau=1,\cdots,T$
  \item for $i=1,\cdots,n$
  \item Sample $\phi_i^{\tau+1}\approx p(\phi_i | \phi_1^{\tau+1},\cdots,\phi_{i-1}^{\tau+1},\phi_{i+1}^{\tau},\cdots,\phi_{n}^{\tau}, Y)$
  \item Iterate over $i$ and $\tau$
  \end{enumerate*}
In this process, the nested iteration assigns a value to each random variable by conditioning it on the current values of the remaining random variables and the observation set. The full execution of this inner loop computes a point in the $n$-dimensional space of the joint distribution $p(\boldsymbol{\phi})$. The outer loop repeats this same process $T$ times to sample from the $n$-dimensional space repeatedly till the joint distribution converges.  
  It has been shown~\cite{gibbs} that the joint distribution $(\phi_1^{\tau},\cdots,\phi_n^{\tau})$ converges geometrically to $p(\phi_1,\cdots,\phi_n|Y)$ as $T\rightarrow\infty$, and therein lies the power of GS.
  \subsection{Trajectory inference through Gibbs sampling}
  \label{sec:gibbs_adaptation}
  In our problem, $Y$ corresponds to the observations $O$, and, $\boldsymbol{\phi}$ corresponds to the uncertain trajectory $U$. The edges in $U$ correspond to the random variables. For a given $O$, the random variables can be identified by taking the union of edges in all paths that connect the nodes in $O$. Now, to employ GS in our problem, we first need to initialize the random variables. This can be achieved by setting each edge in $U$ as either traversed or non-traversed. The conditionals can be computed as outlined in Eq.~\ref{eq:conditional}. However, one key difference from normal GS is that a trajectory is an ordered sequence of random variables. More precisely, the ordering at which each edge is set to be traversed is important and this ordering is used by the recency factor in Eq.~\ref{eq:conditional}. Thus, we need to maintain an additional variable to track the timestamp. Timestamp is set to $0$ at the start of each iteration of the outer loop over $\tau$. For each edge set as traversed in the inner loop, timestamp is incremented by $1$.

  The above steps complete the adaptation of GS for the trajectory inference problem. However, the following two aspects of the algorithm affect its scalability.

  $\bullet$ \textbf{Identifying random variables: } The process to identify the random variables (or edges) by computing all paths connecting the observations is expensive. 

  $\bullet$  \textbf{Computing conditionals:} Computing the conditionals is expensive since we need to scan the entire historical database each time Eq.~\ref{eq:conditional} is computed. Furthermore, this operation happens repeatedly till convergence of the GS. 

 The proposed algorithm, \emph{InferTra}, removes the scalability bottlenecks while maintaining high accuracy. 
    
    %In InferTra, $\mathbb{H}$ is used to learn the spatio-temporal mobility patterns in the form of a \emph{Network Mobility Model (NMM)}. The NMM to compute the conditionals. We detail the algorithm in the next section.
%\section{I\MakeLowercase{nfer}T\MakeLowercase{ra}}
\section{InferTra}
\label{sec:inference}
 InferTra operates in two phases: an offline learning phase to build a Network Mobility Model (NMM), and an online inferencing algorithm using the NMM.
\subsection{Learning a Network Mobility Model}
\label{sec:nmm}
NMM  is a generative model for trajectories and its task is to connect a set of observations (or nodes) without compromising on the mobility patterns of the historical trajectories. The set of all possible paths between a source node and a destination in a road network can be huge. In reality, however, vehicles show affinity towards a limited set of roads that form majority of the trajectories. 
%This skewed distribution results from various factors such as road centrality, speed restrictions, and the inherent human preference towards following their peers, alternatively known as the herd mentality. In addition, preference towards a road segment is often periodic in nature based on the time of the day. For example, shorter but congested roads are  preferred on mornings, but typically avoided during peak hours. 
These spatial and temporal patterns provide a rich characterization of trajectory movements, which InferTra learns through the NMM using a \emph{higher-order Markov Model}. 

The NMM is learned from two sources of input data: a road network $G(V,E)$, and a database of historical trajectories $\mathbb{H}$. In a Markov process, elements of the system make transitions from one state to another based on the preceding history. The length of the history, which dictates the state transitions, is known as the ``order'' of the model. For example, in the most commonly used 1st-order Markov process, the state transition of an element is influenced only by its current state. In the NMM, the state space is defined by nodes $V$ of $G$, and transitions take place only through edges. The transition probabilities are learned from the database of historical trajectories.
The NMM's goal is to better understand the likelihood of a road segment being traversed based on the recent history of a vehicle and the time of the day. These segment traversal likelihoods are modeled as state transitions. Towards that end, we define the concept of an $m$-history sequence of a node $v\in V$, where $m$ is the order of the Markovian process in the NMM and corresponds to the ``recency'' factor in Eq.~\ref{eq:conditional}.
\begin{defn}\textit{
    The $m$-history sequence of a node $v$ is a path $H=\{v_1,\cdots,v_p\}$ through $p$ connected nodes of $G$ such that $p\leq m$ and $v_p=v$.}
\end{defn}
\begin{example}\textit{
    For the road network in Fig.~\ref{fig:motivation}, node `c' has three 2-history sequences: $b\rightarrow c$, $d\rightarrow c$, and $c$.}
\end{example}
 
\begin{defn}\textit{
    \textsc{Spatio-temporal containment.} $H$ is said to be contained in trajectory $T$ at time $t$, if  $\forall i,\;1\leq i\leq |H|,\; H.v_i=T.s_{i+a}.v$, where $\exists a, 0\leq a\leq (|T|-|H|)$. Additionally, $|t-T.s_{(|H|+a)}.t|\leq \delta$. This relationship is denoted using $H\in_t T$.}
\end{defn}

More simply, $H\in_t T$ if $H$ is a sub-sequence of $T$ and the final destination in $H$ is reached within $\delta$ time units from $t$. Incorporating the time of the day in the model allows us to capture the inherent periodicities in mobility patterns. $\delta$ is a user provided parameter. We discuss the semantics of $\delta$ and how to select it below. Prior to that, we introduce the concept of \emph{affinity} toward an edge $e$ based on an $m$-history sequence $H$ of $e.p_1$ . 
\begin{defn}
  \textsc{Edge Affinity.}\textit{
  The affinity $\alpha(e,H,t)$ of an edge $e$ at time $t$ with respect to an $m$-history $H$ of $e.p_1$ is the probability of $e$ being traversed by a trajectory $T$, given that $H\in_t T$. Formally,}

  {\footnotesize
	  \begin{alignat}{2}
		  \nonumber
		  \alpha(e,H,t)=max\{P(H\cup\{e.p_2\}\in_t T|H\in_t T),\epsilon\}&\\
		  \nonumber
	=max\left\{\frac{|\{H\cup\{e.p_2\}\in_t T\;|\;T\in\mathbb{H}\}|}{|\{\forall e'\in E,\; e'.p_1=e.p_1,\;H\cup\{e'\}\in T\;|\;T\in\mathbb{H}\}|},\epsilon\right\}&
      \end{alignat}}
      \textit{where $\epsilon\approx 0$.}
\end{defn}

\begin{figure}[t]
	\centering
		\includegraphics[width=2.6in]{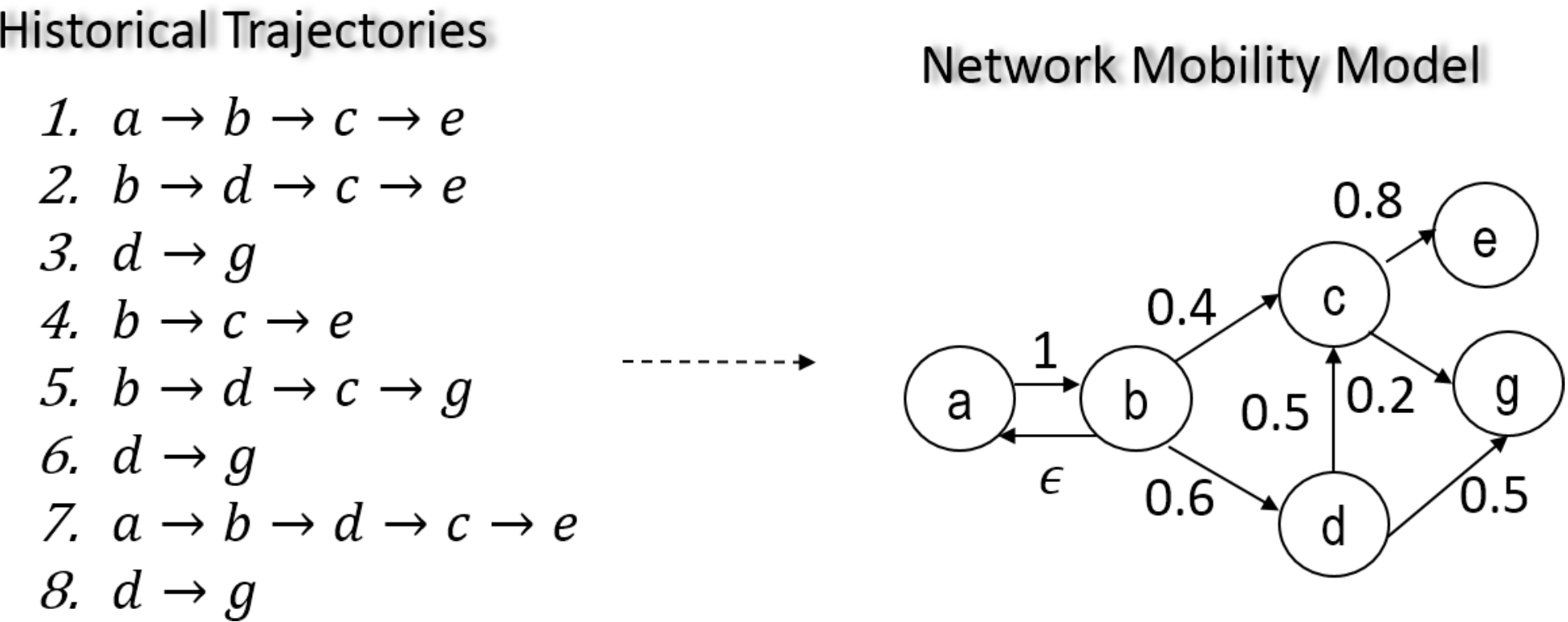}
		\caption{The NMM for the shown trajectories constrained within the road network in Fig.~\ref{fig:motivation}.}
	\label{fig:nmm}
\end{figure}

In essence, $\alpha(e,H,t)$ quantifies the transition probability from $e.p_1$ to $e.p_2$ based on the $m$-history $H$ and timestamp $t$. The NMM is constructed by following this procedure. More specifically, a time window of $\delta$ is slid across all edges, and the corresponding affinities are computed. The distribution of affinities across timestamps is then stored at each edge. Generally, a window size in the range of $\delta=[30,45]$ minutes produces consistent results.
\begin{example}\textit{
    Fig.~\ref{fig:nmm} demonstrates the  NMM for the shown historical trajectories constrained within the network in Fig.~\ref{fig:motivation}. We assume $m=1$ and all segments are traversed at the same timestamp for simplicity, . We thus ignore the temporal aspect. }
\end{example}

While the time of the day certainly influences the mobility pattern, storing the entire affinity distribution for each edge and $m$-history pair incurs a large storage cost. For instance, sliding a time window of $30$ minutes would produce $24*60-30=1410$ affinity values for each pair. For a vast majority of the segments, the affinities remain constant with time, and storing the entire distribution promotes redundancy. To remove this redundancy, we partition these distributions into the optimum number of bins using the \emph{Freedman-Diaconis} rule~\cite{freedman}. The Freedman-Diaconis rule states that the width $w$ of each bin in the distribution $\alpha(e,t)$ should be:
\begin{equation}
  w=2\frac{IQR(\alpha(e,H,t))}{n^{\frac{1}{3}}}
\end{equation}
where $n$ is the number of time windows, and $IQR(\alpha(e,H,t))$ denotes the \emph{interquartile range} of the affinity distribution at $e$ for $m$-history $H$. The interquartile range is a measure of the statistical dispersion of a distribution and is equal to the difference between the third and the first quartile, i.e., the 75th and 25th percentile of $\alpha(e,H,t)$. Generally, the Freedman-Diaconis rule is based on minimizing the sum of the squared errors between the bin height and the underlying actual distribution. Based on this rule, the number of bins at each edge is automatically learned and set to $\frac{n}{w}$.
%\subsubsection{Parameters}

\textbf{Selecting $\boldmath{m}$: }
While a longer history takes a global view, it creates an explosion in storage cost. 
%Additionally, as discussed in Sec.~\ref{sec:gibbs_adaptation}, estimation of transition probabilities gets inaccurate due to sample space sparsity as $m$ increases. %On the other hand, a small $m$ relies on locally optimal decisions. %Intuitively, most vehicles take locally optimal decisions and thus a history beyond a certain $m$ does not provide any significant boost to accuracy. 
We optimize $m$ by analyzing the storage vs.  accuracy trade off in training data. 
%As shown in Sec.~\ref{sec:experiments}, the accuracy saturates at $m=3$.
%The NMM makes an underlying Markovian assumption whereby, the affinity at any edge is dependent only on the current location of a trjectory. Theoretically, it is indeed possible that the entire sequence of past locations
\subsection{Inferencing from the NMM}
\label{sec:inferrence}
%The NMM provides the platform for inferring uncertain trajectories from partial observations. In this section, we develop the inference phase of the InferTra algorithm.	
The NMM models the likelihood of a segment being traversed based on its $m$-history and the current time of a day. By performing \emph{semi-supervised random walks with restarts} on the NMM, we compute the conditional of an edge being traversed without performing any trajectory scans. Using the principle of GS, by repeatedly sampling from these conditionals, we compute the conditional joint distribution $p(U|O)$.
%More specifically, given a sequence of partial observations $O=\{s_1,\cdots,s_n\}$, we pick each pair of consecutive observations $s_i$ and $s_{i+1}$ and generate random trajectories from the NMM that connects $s_i$ to $s_{i+1}$. The likelihood of a trajectory being generated from NMM conforms to the patterns displayed by the historical trajectories and therefore captures an accurate reflection of the possibilities. We generate these random trajectories by performing \emph{semi-supervised random walk with restarts} on the NMM.
\subsubsection{Semi-supervised Random Walk with Restarts}
\label{sec:rwr}
Random walk with Restarts (RWR) simulate the trajectory of a random walker who starts from a source node and iteratively jumps from one node to a neighbor. The probability of jumping to a neighbor is proportional to the weight of the corresponding edge. At the same time, with a restart probability $r$, the walker jumps back to the source node.  Upon returning to the source node, a new walk is initiated.

\textbf{Trajectory Generation: }We generate trajectories from the NMM through semi-supervised RWR.  Each generated trajectory represents a sample from the joint distribution space of $p(U|O)$. By repeatedly generating these trajectories, the uncertain trajectory $U$ is inferred.

Given a set of observations, each pair of consecutive observations, $s_i$ and $s_{i+1}$, is picked and a RWR is initiated from $s_i.v$ (Alg.~\ref{alg:sampletraj}). As in a normal RWR, for each jump, the new destination is selected based on the affinities of all outgoing edges from $s_i.v$. However, to model the mobility pattern in a trajectory, we only consider edges that do not create a cycle (line 11 in Alg.~\ref{alg:sampletraj}). This follows from the underlying assumption that each transition in a trajectory takes us closer to the destination. To avoid cycles, we maintain a set $\mathcal{S}$ that stores each visited edge while performing the walk. Now, given a current node $v$ and its $m$-history at time $t$, the chances of selecting an outgoing edge $e$ is proportional to its affinity (line 13 in Alg.~\ref{alg:sampletraj}). Mathematically, the transition probability through an edge $e$ is expressed as:
\begin{equation}
  \label{eq:transitionprobability}
   p(e)=\begin{cases}
     (1-r)\frac{\alpha(e,H,t)}{\sum_{\forall e' \in \bar{E}}\alpha(e',H,t)} &\mbox{if $e\in \bar{E}$}\\
     0&\mbox{otherwise}
\end{cases}
\end{equation}

$\bar{E}=\{e.p_1=v, e.p_2\not\in \{e'.p_1 \cup e'.p_2 | \forall e' \in \mathcal{S}\}\;|\;e\in E\}$ is the set of outgoing edges that do not induce a cycle. $v$ is the current node, and $t$ is the current time (line 14 in Alg.~\ref{alg:sampletraj}).

\textbf{Capturing destination bias: }An edge transition to $e$ in the random walk does not compute $e$'s conditional probability. As formulated in Eq.~\ref{eq:conditional}, the conditional probability is equivalent to computing the transition probability on the subset of historical trajectories that pass through the destination $s_{2}$. Computing Eq.~\ref{eq:conditional} from $\mathbb{H}$ requires us to perform trajectory scans, which, as we have shown in Sec.~\ref{sec:gibbs_adaptation} is extremely expensive. Thus, to approximate Eq.~\ref{eq:conditional} in a scalable manner, we enforce the destination bias on RWR itself using the restart probability $r$. In Eq.~\ref{eq:transitionprobability}, only $(1-r)$ of the probability mass is distributed among edge transitions. Like in a normal RWR, there is a chance of jumping back to the source node with probability $r$. If a restart occurs, semantically, the walk is considered to have ventured towards a wrong direction and thus discarded by reinitializing $\mathcal{S}$ to an empty set (line 9 in Alg.~\ref{alg:sampletraj}). A new walk is then initiated with the goal of reaching destination $s_{i+1}.v$, and finally, only those walks that successfully reach $s_{i+1}.v$ are recorded. Consequently, the destinations bias is strictly enforced on all edge transitions.
%in Sec.~\ref{sec:restartprobability}.  

\begin{algorithm}[t]
	%\caption{SampleTrajectory($\mathcal{P}$, $s_1$, $s_2$)}
	\caption{SampleTrajectory($s_1$, $s_2$)}
\label{alg:sampletraj}
{\scriptsize
\begin{algorithmic}[1]
%\REQUIRE {$\mathbb{N}=\{N_i=(V_N,E_N,L_i,S_i)\}$ is the dynamic network}
%\REQUIRE{$f(g)$ is a user-provided monotonic function}
%\REQUIRE{$\theta$ is the threshold for defining $\phi(G)$}
%\REQUIRE{$k$ is the user-provided size of the answer set}
%\ENSURE {$\mathbb{A}$ is the greedy answer set}
  \STATE {$curr\leftarrow s_1.v$}
  \STATE {$t\leftarrow s_1.t$}
  \STATE {$r\leftarrow$ restart probability selected based on Eq.~\ref{eq:restartprobability}}
  \STATE {$\mathcal{S}\leftarrow \emptyset$}
  \WHILE {$curr \neq s_2.v$}
  \STATE $p \leftarrow \mbox{sample uniformly from }[0,1]$
  \IF {$p\leq r$}
  \STATE{$curr\leftarrow s_1.v$}
  \STATE {$\mathcal{S}\leftarrow \emptyset$}
  \ELSE
  \STATE {$\bar{E}\leftarrow \{e.p_1=curr,\;\mbox{$e\in E$ does not create a cycle in current walk} \}$}
  \STATE {$H\leftarrow$ extract $m$-history of $curr$ from $\mathcal{S}$}
  \STATE $e\leftarrow$ select edge from $\bar{E}$ proportional to $\frac{\alpha(e,H,t)}{\sum_{\forall e' \in \bar{E}}\alpha(e',H,t)}$
  \STATE {$t\leftarrow t+speed(e,t)*length(e)$}
  \STATE {$curr\leftarrow e.p_2$}
  \IF {$curr=s_2.v$ \textbf{and} $p\leq \tau$}
  \STATE{$curr\leftarrow s_1.v$}
  \STATE {$\mathcal{S}\leftarrow \emptyset$}
  \ELSE
  \STATE {$\mathcal{S}\leftarrow\mathcal{S}\cup \{e\}$}
  \ENDIF
  \ENDIF
\ENDWHILE
\STATE \textbf{return} $\mathcal{S}$
\end{algorithmic}}
\end{algorithm}

In a traditional RWR, the restart probability is static since the goal is to compute network proximity to the source node. In our problem, the goal is to reach the destination node $s_{i+1}.v$ within the expected duration $X_t=s_{i+1}.t-s_i.t$. Otherwise, it is likely that the walker is in the wrong path. Therefore, sufficient time must be allowed to the walker to successfully reach the target. However, if the walker is unsuccessful in reaching the destination within $X_t$, then the chances of restarting the walk should be explored. 
%To model these requirements, for a given pair of points $s_i$ and $s_{i+1}$, we first define the \emph{expected time}, $X_t=s_{i+1}.t-s_i.t$. Now, 
To model these requirements, we define the restart probability $r$ as the following.
%we define $t$ be time spent on the ``current'' walk. $t$ is computed from the average speed and length of each traversed edge as shown in line 14 of Alg.~\ref{alg:sampletraj}. Now, intuitively, we should not restart the walk till $t<X_t$. However, as $t$ exceeds $X_t$, aborting the walk should become a possibility. Mathematically, we model the restart probability $r$ as the following.
\begin{equation}
	\label{eq:restartprobability}
	r=1-\frac{1}{e^{\frac{max\{0,t-X_t\}}{X_t}}}
\end{equation}

where  $t$ is the time spent on the ``current'' walk (line 14 of Alg.~\ref{alg:sampletraj}). Thus, till the time spent on a walk is less than the expected time $X_t$, the restart probability is $0$. As $t$ exceeds $X_t$, the restart probability begins to increase exponentially. 

\begin{figure}[t]
	\centering
		\includegraphics[width=2.9in]{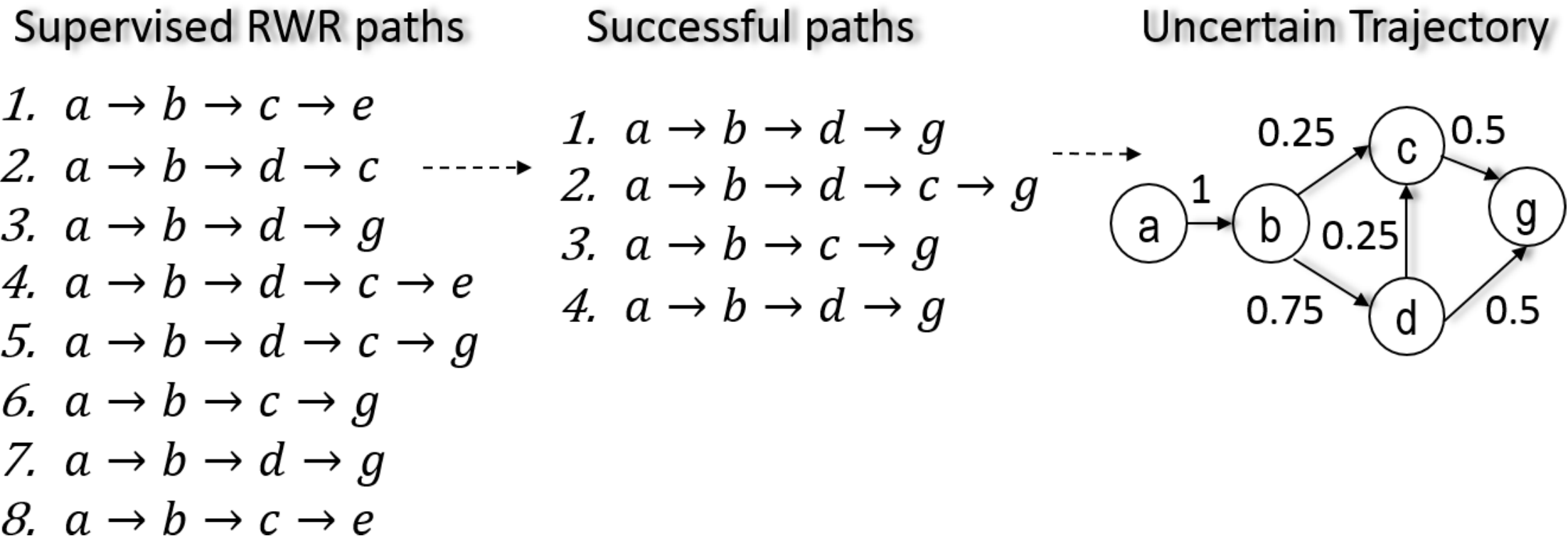}
		\caption{The uncertain trajectory constructed from the NMM in Fig.~\ref{fig:nmm} on the observations $\{a\rightarrow g\}$.}
	\label{fig:rwr}
\end{figure}

To summarize, we generate trajectories by repeatedly sampling from the conditionals expressed through edge transition probabilities and the destination bias. Each generated trajectory is a point in the joint distribution space of $p(U|O)$. Thus, $U$ is defined over all edges that are sampled at least once. As outlined in Eq.~\ref{eq:marginal}, 
\begin{equation}
  edgeWeight(e)=\frac{\|\{T\in\mathbb{T}|e\in T\|}{\|\mathbb{T}\|}
\end{equation}
where $T$ is the trajectories generated through RWR. RWR terminates once the joint distribution of $U$ converges 
\begin{example}
  \textit{ Fig.~\ref{fig:rwr} shows a probable result  with respect to the road network in Fig.~\ref{fig:motivation} and its corresponding NMM in Fig.~\ref{fig:nmm}, under observations $\{a\rightarrow g\}$. From $8$ different walks, $4$ are successful in reaching destination $g$. For simplicity we ignore the temporal aspect and assume that the estimated time of all successful paths match the observed time. In other words, $\tau\approx 0$. From these successful walks, the uncertain trajectory is constructed. 
Note that although $80$\% of the vehicles go to $e$ from $c$ in the NMM, due to the destination bias enforced by $g$, the $c\rightarrow g$ edge has a probability of $50$\% of being traversed. 
}
  \end{example}
\subsection{Properties of Uncertain Trajectories}
\label{sec:properties}
We next highlight the properties of an uncertain trajectory $U(V',E')$.

\textbf{Node Visit Likelihood: }The probability of visiting a node $v \in V'$ is the sum of the incoming edge-weights $E_{in}=\{e.p_2=v| e\in E'\}$. 
\begin{equation}
  \label{eq:nodeprobability}
  nodeWeight(v)=\sum_{\forall e \in E_{in}}edgeWeight(e)
\end{equation}

The sum of incoming edge weights in a node $v$ is equal to the sum of its outgoing edge weights. 
%Furthermore, if $v$ is part of partial observations $O$ from which $U$ is inferred, i.e., $\exists s\in O, s.v=v$, then $nodeWeight(v)=1$.

\textbf{Trajectory Likelihood: }
%The inferred uncertain trajectory captures the entire spectrum of possibilities. Indeed, it allows inference of the maximum likelihood trajectory. 
Let $\mathbb{P}$ be the set of paths from the source to the destination of $U$. The probability of a trajectory $T\in\mathbb{P}$ is expressed as the following.
\begin{equation}
  \label{eq:trajectoryprobability}
  p(T)=\prod_{i=2}^{|T|}p(v_i\in T| v_{i-1}\in T)
\end{equation}

where,
\begin{equation}
  \nonumber
 p(v_i\in T| v_{i-1}\in T)=\frac{edgeWeight((v_{i-1},v_i))}{\sum_{\forall e\in E', e.p_1=v_{i-1}}edgeWeight(e)}
\end{equation}

Taking products of the individual node likelihoods that constitute a trajectory $T\in\mathbb{P}$ is not enough since they are not independent. Rather, it is the probability of reaching node $v_{i}$ given that $v_{i-1}$ has already been reached. 
%Thus, Eq. \ref{eq:trajectoryprobability} computes the likelihood iteratively starting from $v_2$, since $p(v_1)=1$.
\begin{comment}
\begin{example}\textit{
    Revisiting Fig.~\ref{fig:rwr}, let us compute the probability of $T=a\rightarrow b\rightarrow c \rightarrow g$. $p(b\in T| a\in T)=1$. Next, $p(c\in T| b\in T)=\frac{0.25}{0.75+0.25}=0.25$. Finally, $p(g\in T| c\in T)=\frac{0.5}{0.5}=1$. Thus, $p(T=a\rightarrow b\rightarrow c \rightarrow g)=0.25$.}
\end{example}
\end{comment}
The \textit{maximum likelihood trajectory} is, 
 $T^*=\arg\max_{\forall T\in\mathbb{P}}p(T)$.
  \subsection{Discussion}
  \label{sec:discussion}
  This section answers two important questions. \textit{Why RWR on the NMM is equivalent to Gibbs sampling?} And \textit{Why InferTra is more efficient than the straight-forward implementation of Gibbs sampling outlined in Sec.~\ref{sec:gibbs}? }
  
The first step in GS is to identify the vector of random variables and initialize them. The random variables can be initialized to arbitrary values since it is independent from the final convergence to the joint distribution. In InferTra, the first edge transition is conditioned only on the observed data, which translates to initializing all edges as ``not-traversed''.  

After initialization, GS moves into the iterative phase. When compared to the outline in Sec.~\ref{sec:gibbs}, in RWR, each new walk corresponds to the outer loop over $\tau$, and the edge transitions correspond to the inner loop. Let us first focus on the inner loop, which samples each edge conditioned on the observations and the current values of remaining edges. In GS the inner loop is iterated over all edges. In RWR, the sampling stops as soon as the walk reaches the destination. This early termination, however, does not cause any loss of information, as stated in Theorem~\ref{thm:walk1}.
\begin{theorem}
  \label{thm:walk1}\textit{
  Let $i=m$, when the random walker reaches the destination. For all subsequent iterations of $i$}
\begin{equation}
  p(e_i|e_1^{\tau+1},\cdots,e_{i-1}^{\tau+1},e_{i+1}^{\tau},\cdots,e_n^{\tau},O)=0 
\end{equation}
\end{theorem}
\textsc{Proof:} Recall, that all trajectories are cycle-free. Thus, for all subsequent iterations or $i$, the probability of transitioning to any other edge and returning to the destination is $0$.$\hfill\square$

The second key difference from GS is that each new iteration over $\tau$ does not forget the values of the random variables assigned in the previous iteration. More specifically, the conditional at $i=1$ at any iteration of $\tau$ is expressed as following:
\begin{equation}
  e_1^{\tau+1}\approx p(e_1|e_2^{\tau},\cdots,e_{n}^{\tau},O) 
\end{equation}
In a new RWR walk, the previous walk is completely forgotten and the first transition to edge $e_1$ is conditioned only on the observations. This deviation, however, does not cause any information loss as stated in the following theorem.
\begin{theorem}
  \label{thm:walk2}
  \begin{alignat}{2}
    \nonumber
  p(e_i|e_1^{\tau+1},\cdots,e_{i-1}^{\tau+1},e_{i+1}^{\tau},\cdots,e_n^{\tau},O)&=\\
  \hspace{-2in}p(e_i|e_1^{\tau+1},\cdots,e_{i-1}^{\tau+1},O)
\end{alignat}
\end{theorem}

\textsc{Proof:} From Eq.~\ref{eq:conditional}, we know that the conditional depends only on the past $m$ traversed edges and the observation set. Now, as discussed in Sec.~\ref{sec:gibbs_adaptation}, since trajectory is a sequence of random variables, for each edge that is marked as traversed, a current timestamp value is also assigned to it. Furthermore, the timestamp is reset to $0$ at the start of each iteration of the outer loop. Let us denote the set of traversed edges at the current iteration of $\tau$ as $\mathbb{E}$. It is easy to see that $\mathbb{E}\subseteq \{e_1,\cdots,e_{i-1}\}$, where $i$ is the latest edge being sampled in the inner loop. Thus, edges sampled in previous iterations of $\tau$ have no impact in the current iteration, which is how the RWR operates. $\hfill\square$

Above analysis establishes how RWR on NMM conforms to GS framework. Now, we focus on the efficiency of InferTra. The naive pipeline of Sec.~\ref{sec:gibbs_adaptation} suffers from two scalability issues. First issue is of identifying the random variables, which finds all paths connecting each pair of consecutive observed nodes. RWR completely skips this step. Since in the initialization step, we initialize all edges as not traversed, we do not need to identify these edges explicitly. We identify these edges on-the-fly during the random walk. 

The second aspect affecting scalability of GS is computing the conditionals. This step is expensive since computing the conditionals requires scanning the entire trajectory database. InferTra tackles this problem by completely negating the need to perform database scans in the online phase. The NMM pre-computes the $m$-history for each transition in its offline learning. To condition the transition probabilities with the destination bias as required by Eq.~\ref{eq:conditional}, RWR only records those walks that successfully reach the destination. Furthermore, using Theorem~\ref{thm:walk1}, the number of conditionals computed is restricted to only those that have non-zero likelihoods. 
\section{Experiments}
\label{sec:experiments}
In this section, we show that:
\begin{itemize*}
	      \item \textbf{Inference:} InferTra is more accurate and scalable than the state-of-the-art trajectory inferencing technique.
	      \item \textbf{Versatility: } InferTra supports a wider range of queries.
\end{itemize*}
\subsection{Datasets}
\label{sec:datasets}
\textbf{GPS traces: }We use gps-traces of cabs in the city of Beijing~\cite{cab1,cab2}. Each cab is tracked for a week-long duration. Prediction on this dataset is particularly difficult, since cabs do not have any common or frequent routes that are typically observed in trajectories of buses or personal vehicles.

\textbf{Road network: }The road network of Beijing is extracted from OpenStreetMap~\cite{osm}. The Beijing road network contains 623,975 nodes and 672,284 edges. 

\textbf{Network-constrained trajectories: } The trajectories are map-matched~\cite{map1,map2,map3} to the Beijing network generating $136,759$ network-constrained trajectories.  
\subsection{Experimental Setup}
\label{sec:setup}
Our algorithms are implemented in Java JDK 1.6.0 and evaluated on a PC with 12GB memory and Intel i5 2.60GHz quad core processor running Ubuntu 13.04. We benchmark InferTra against HRIS~\cite{hris}, the shortest path (SP), and the shortest time path (STP) in the road network. To compute STP, we use average traversal times in the historical data as edge weights. 

%\subsubsection{Parameters}
\textbf{Parameters: }
 HRIS contains $8$ different parameters, which are set as suggested by the authors in~\cite{hris}. InferTra has only two parameters: the sliding time window size $\delta$, and the model order. $\delta$ is set to $30$. The order of the Markov model for NMM is learned from the training dataset and is set to $3$. The learning procedure is discussed in Sec.~\ref{sec:infertraperformance}.  The default sampling interval (SI) is assumed to be $15$ minutes per sampled point.

\textbf{Benchmarking Setup}
To evaluate prediction accuracies, we perform $10$-fold cross validation. The training dataset is used to learn the NMM and inference is performed on the test set. To model a desired sampling rate, nodes from trajectories in the test set are deleted accordingly. Next, a prediction $U_p=(V_p,E_p)$ is generated on the under-sampled trajectory and compared with the original ground truth trajectory $T=(V,E)$. For InferTra, $U_p$ is an edge-weighted graph, whereas for HRIS/SP/STP, $U_p$ is a path in the road network. The accuracy of the prediction is quantified using \emph{F-score}. F-score can be visualized as a weighted average of the precision and recall, where the best performance corresponds to a value of $1$, and the worst corresponds to $0$. 
For HRIS (or SP/STP), computing precision and recall is straightforward. In an uncertain trajectory however, a constituent edge exists with a probability. Thus, the formulations of precision and recall are modified to handle both certain and uncertain trajectories.
\begin{comment}
\begin{equation}
	\mbox{F-score}&=2\:\frac{precision\times recall}{precision + recall}	
\end{equation}
\end{comment}
%Next, we explain how precision and recall is computed on the predictions. 
Let $E_c=E\cap E_p$ be the set of common edges in $T$ and $U_p$. Now,
\begin{alignat}{2}
	\label{eq:recall}
	\mbox{recall}&=\frac{\sum_{\forall e \in E_c}edgeWeight(e)}{|E|}\\
	\label{eq:precision}
	\mbox{precision}&=\frac{\sum_{\forall e \in E_c}edgeWeight(e)}{\sum_{\forall e \in E_p}edgeWeight(e)}
\end{alignat}
\begin{figure}[t]
	\centering
		\includegraphics[width=3.3in]{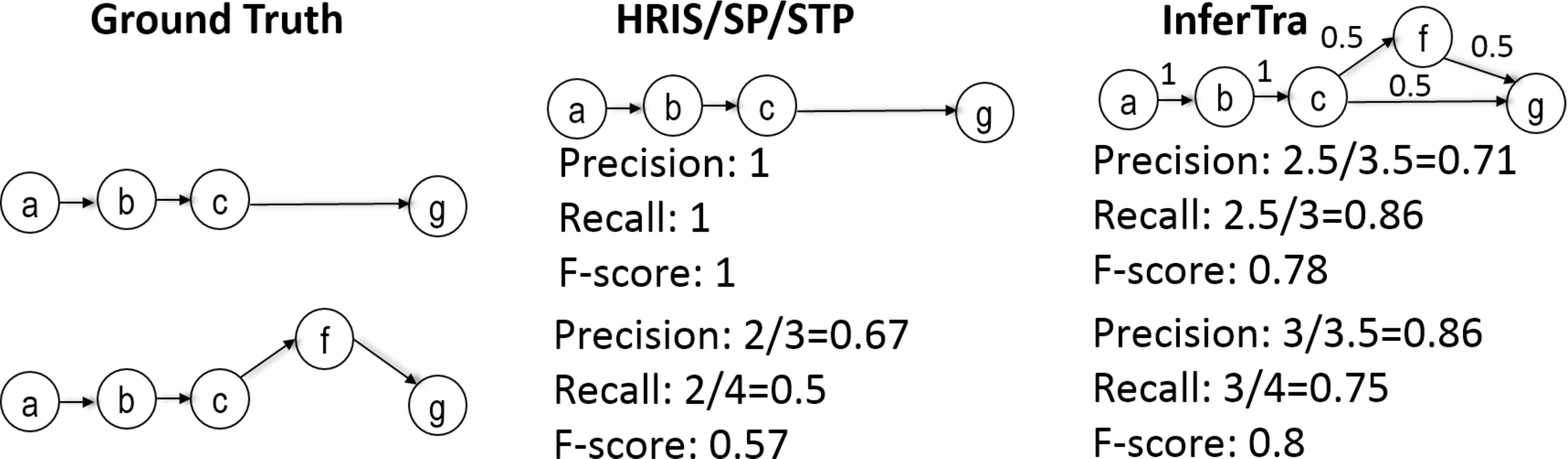}
		\caption{Precision, Recall and F-score values for HRIS/SP and InferTra against the two shown ground truth trajectories.}
	\label{fig:fscore}
\end{figure}
Eqs. \ref{eq:recall} and \ref{eq:precision} degenerate to their standard formulations for certain trajectories. 
For uncertain trajectories, rather than operating in a binary world, their likelihoods are considered. 
\begin{example}
  \textit{	Fig. \ref{fig:fscore} demonstrates the F-score computations. }
  %F-score quantifies how well the constituent edges of the ground truth have been captured in the prediction. Note that taking edge likelihood products of the prediction is not enough, since even a single missing edge would produce a likelihood of 0. For instance, the likelihood of HRIS/SP/STP on the second ground-truth would be $0$, even though it correctly predicts the presence of the sequence $a\rightarrow b\rightarrow c$. For this same reason, HRIS\cite{hris} also computes the minimum of precision and recall to evaluate their predictions.}
\end{example}
\subsection{Performance of naive Gibbs sampling}
\label{sec:gibbsperformance}
Before benchmarking the performance of InferTra, we revisit the performance issues of GS highlighted in Sec.~\ref{sec:gibbs_adaptation}. 
%The non-scalability stems from two steps: identifying the random variables and computing conditionals. 

\begin{figure}
	\centering
		\subfigure[Identifying random variables]{
	\label{fig:samplingRateVsrandomvariables}
		\includegraphics[width=1.50in]{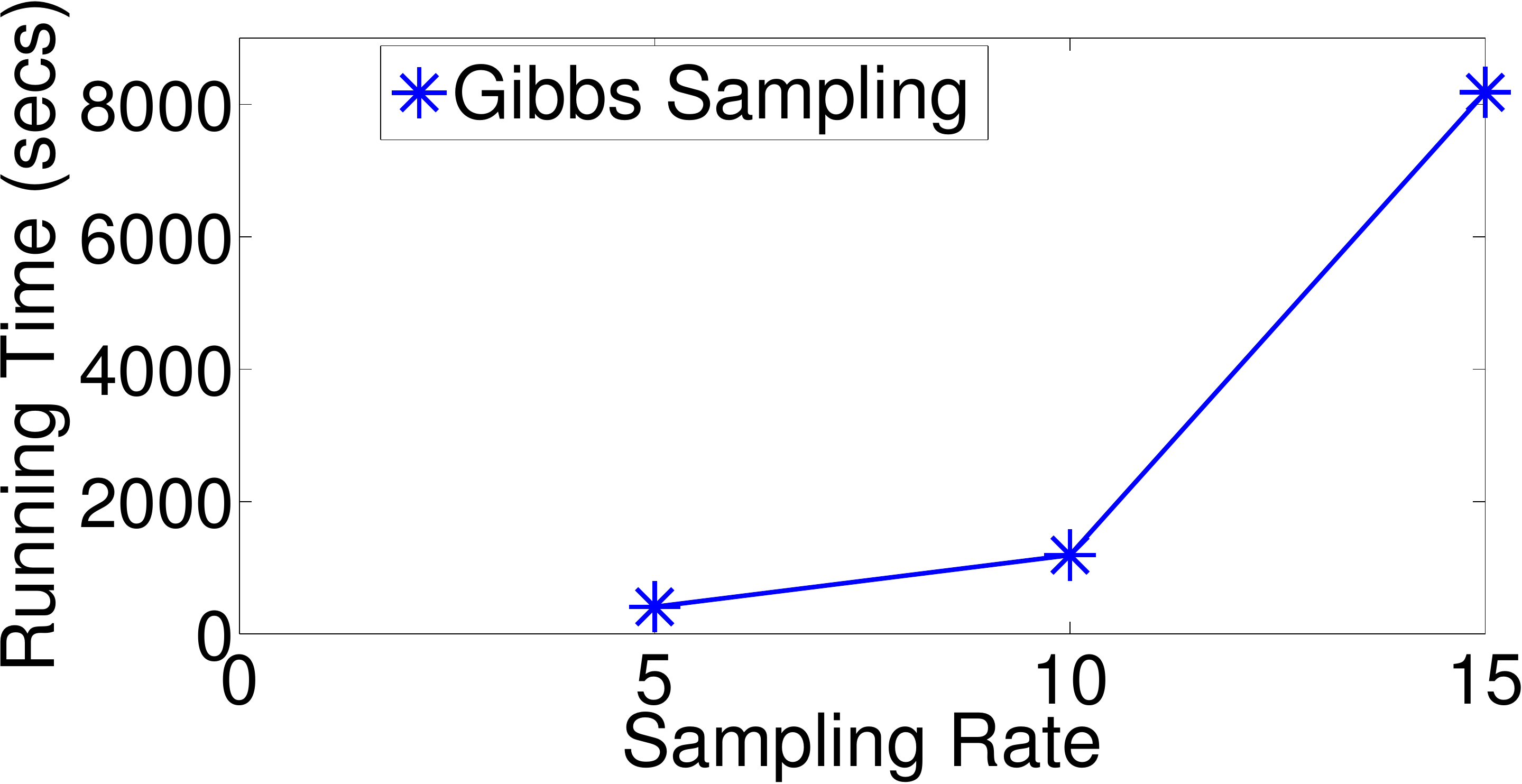}}
		\subfigure[Computing Conditionals]{
	\label{fig:samplingRateVsconditional}
		\includegraphics[width=1.50in]{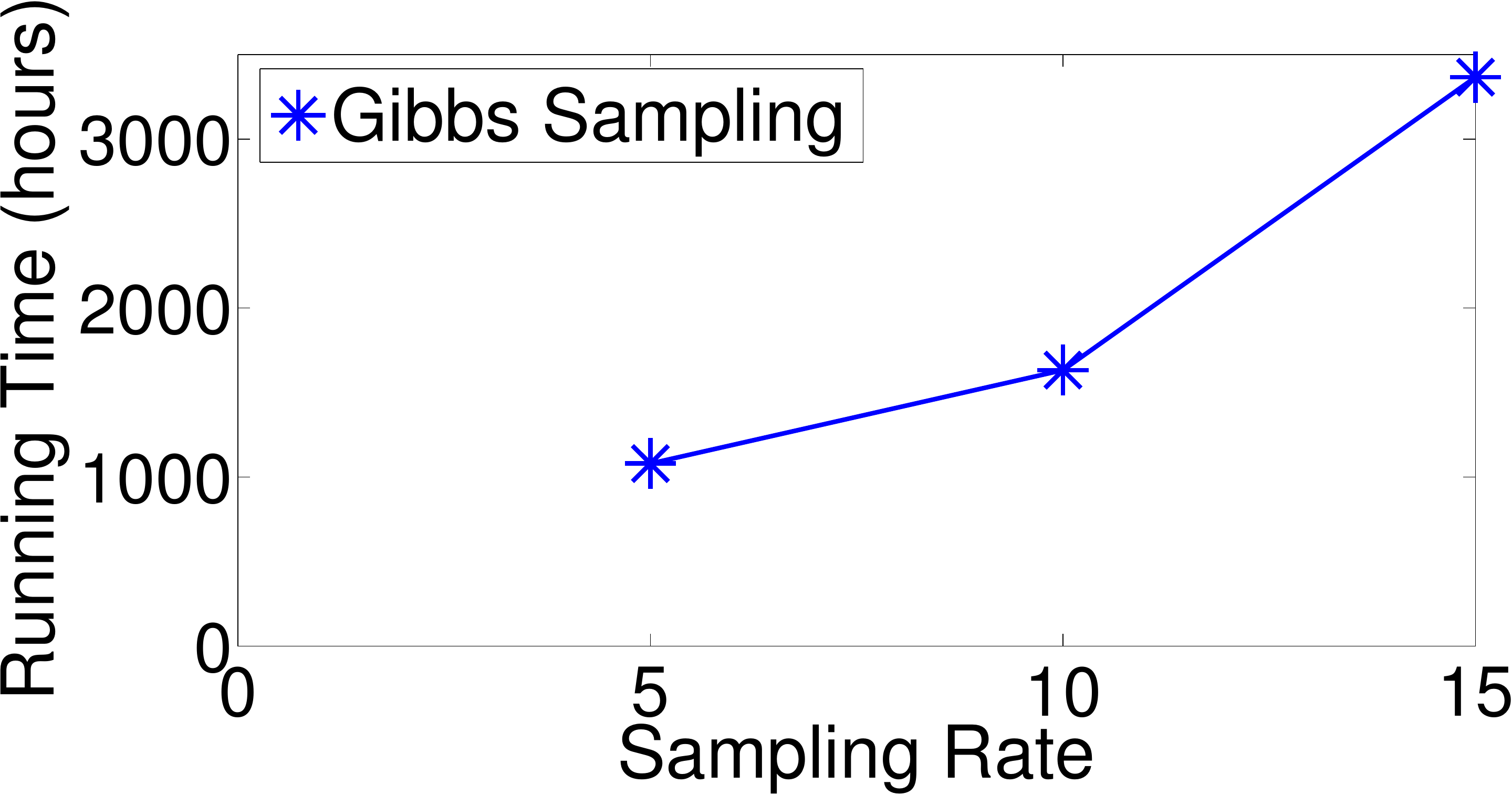}}
\caption{Growth rate of running time with sampling interval (SI) while (a) Identifying random variables, (b) Computing conditionals.}
\end{figure}

  $\bullet$ \textbf{Identifying random variables: }Fig.~\ref{fig:samplingRateVsrandomvariables} shows the growth rate of time taken to identify the random variables with SI. To keep the path identification practical, we restrict ourselves to only those paths that are at most $2.5$ times the length of the shortest path. Even then, at $SI=15$, it takes $8183$ seconds to identify all paths. 
  %Given that this operation needs to be performed for each pair of consecutive observations, the overall running time of gets prohibitively large. 

  $\bullet$  \textbf{Computing conditionals:} Computing the conditionals presents an even larger scalability challenge. At $m=3$, computing Eq.~\ref{eq:conditional} takes $1.2$ seconds on average. Consequently, even at an SI of $5$, it is practically infeasible to achieve convergence. Assuming it takes at least $10000$ iterations of $\tau$ given the high dimension of the space, we compute the average number of random variables, $d$, at SI$=5,10,15$, and compute the projected convergence time using the formula $1.2*10000*d$. Figure \ref{fig:samplingRateVsconditional} demonstrates this projected time. Clearly, $1000$ hours even at SI$=5$ is prohibitively large.
%Similarly, the denominator in precision for HRIS is $|E_p|$, while the numerators are $|E_o|$ in both precision and recall. 
\begin{figure*}[t]
	\centering
		\subfigure[]{
	\label{fig:samplingRateVsAccuracy}
		\includegraphics[width=2.20in]{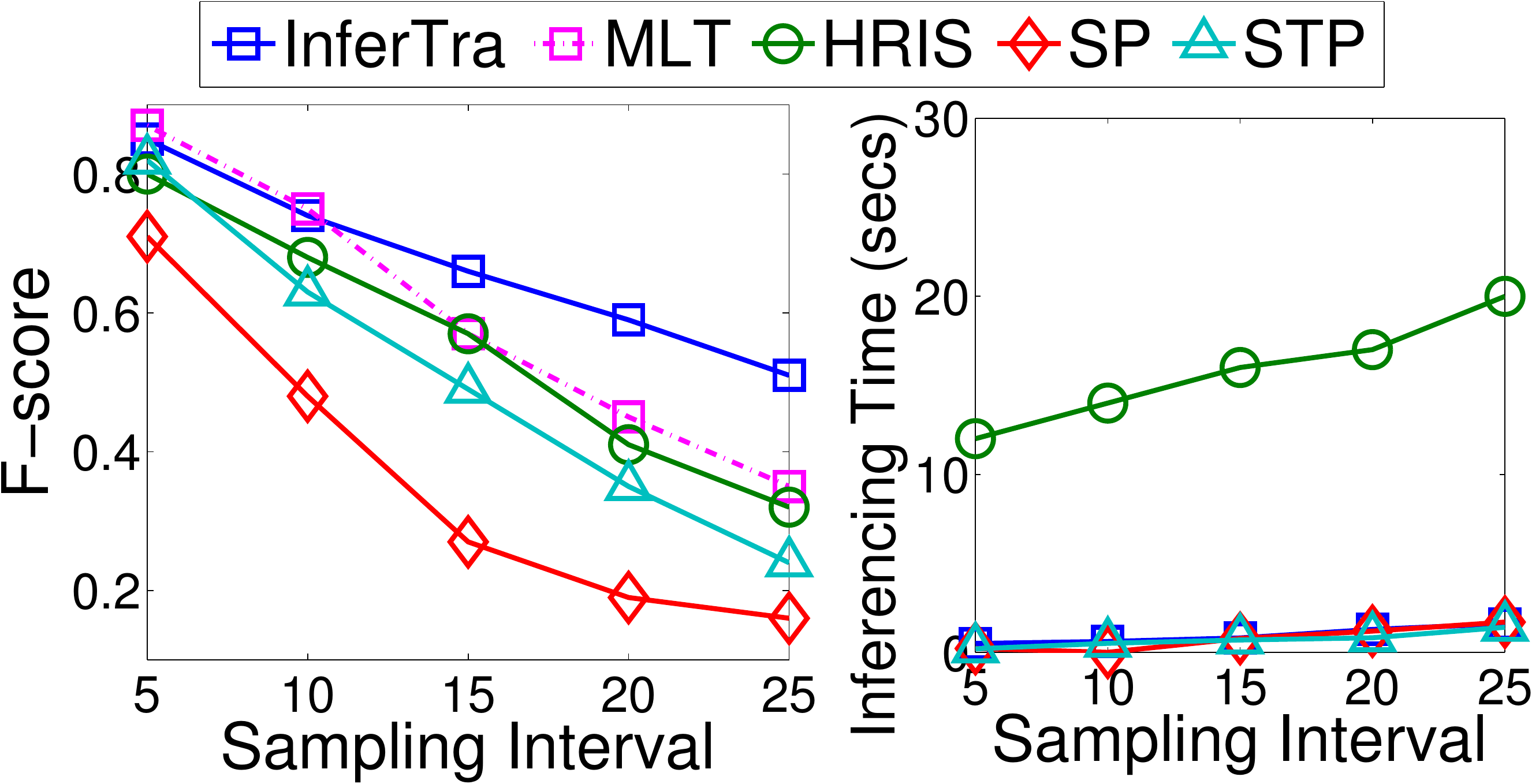}}
		\subfigure[]{
	\label{fig:trainingSizeVsTime}
		\includegraphics[width=1.10in]{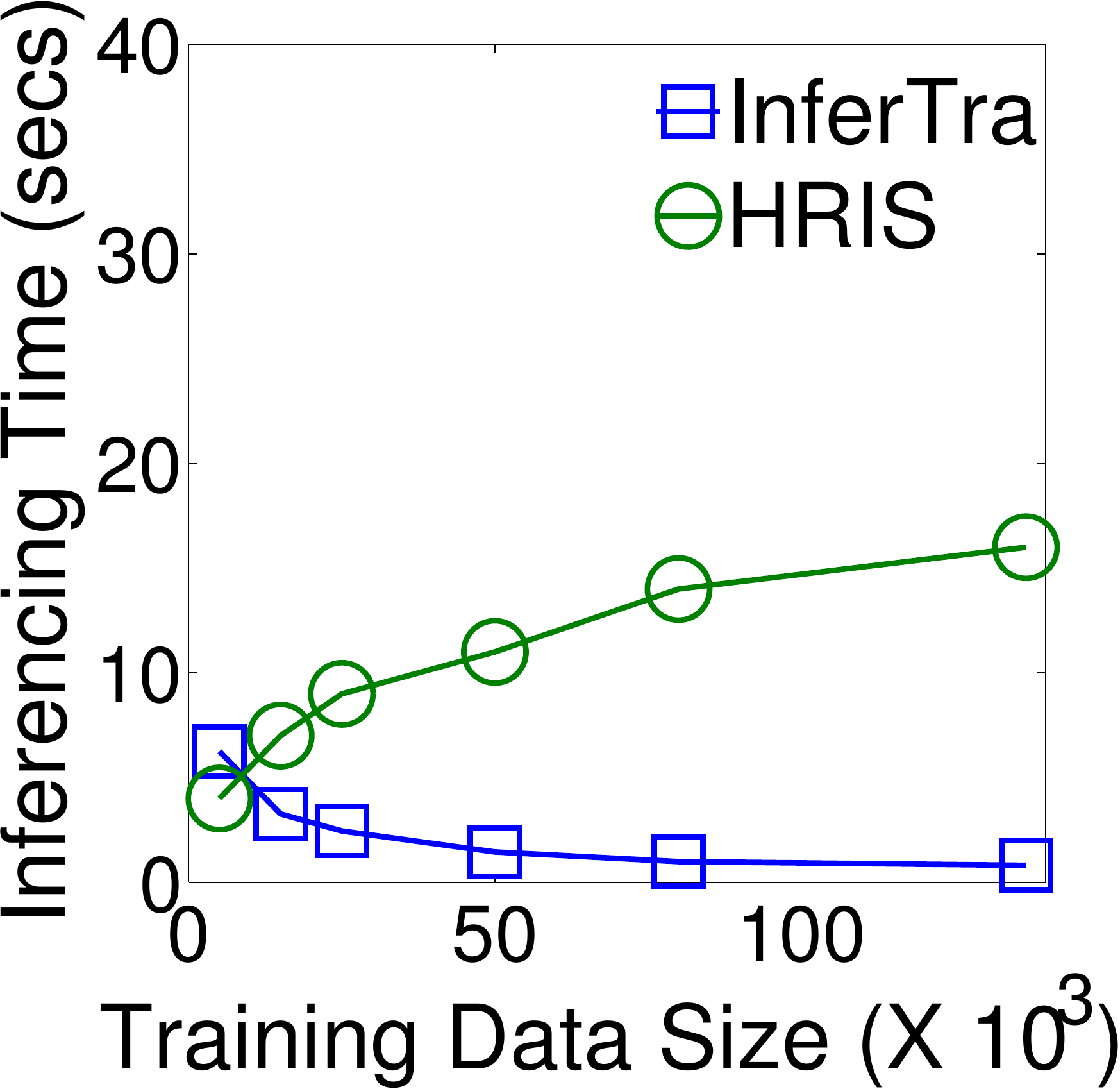}}
		\subfigure[]{
	\label{fig:trainingSizeVsAccuracy}
		\includegraphics[width=1.10in]{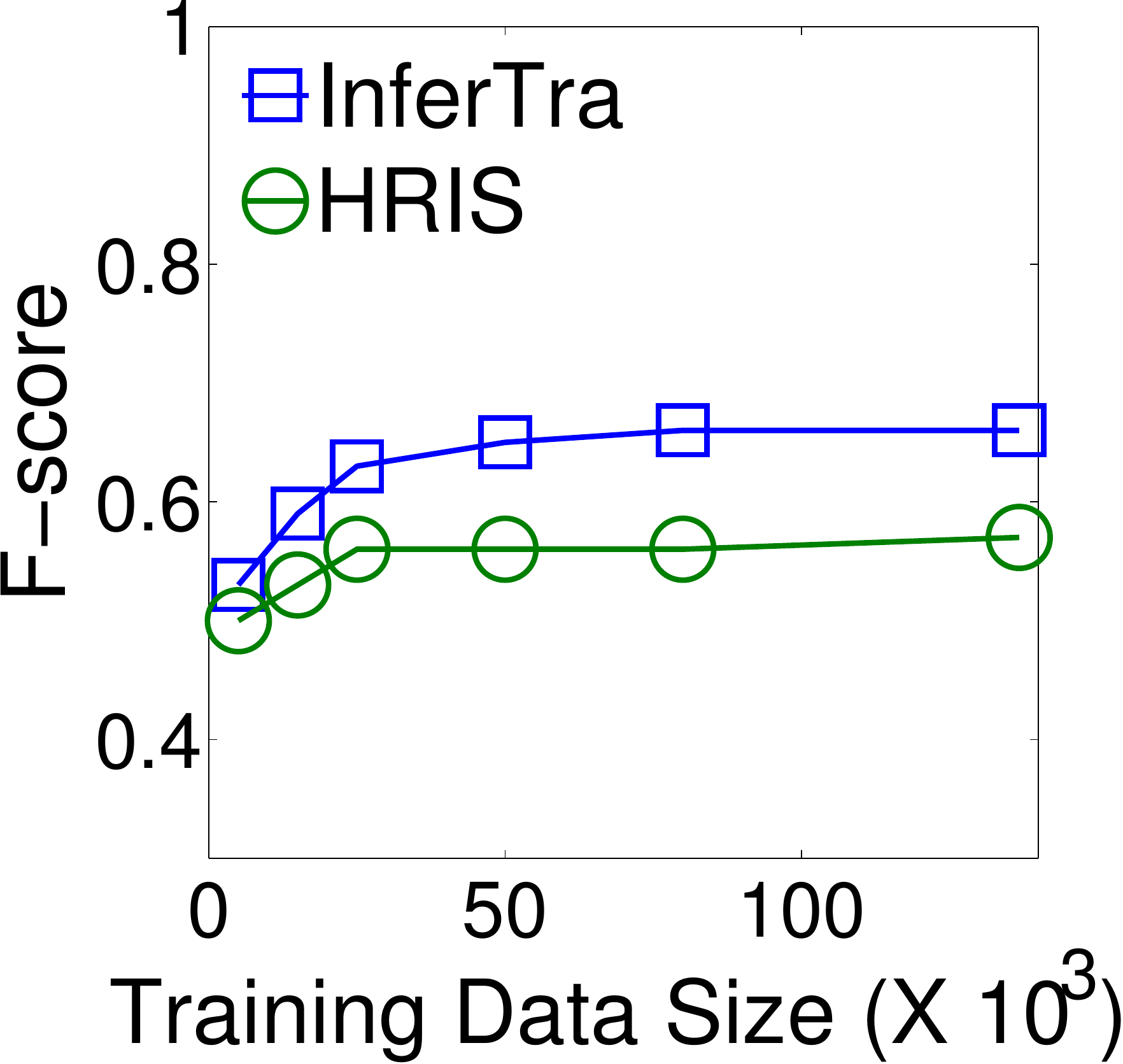}}
		\subfigure[]{
	\label{fig:markovVsAccuracy}
		\includegraphics[width=2.20in]{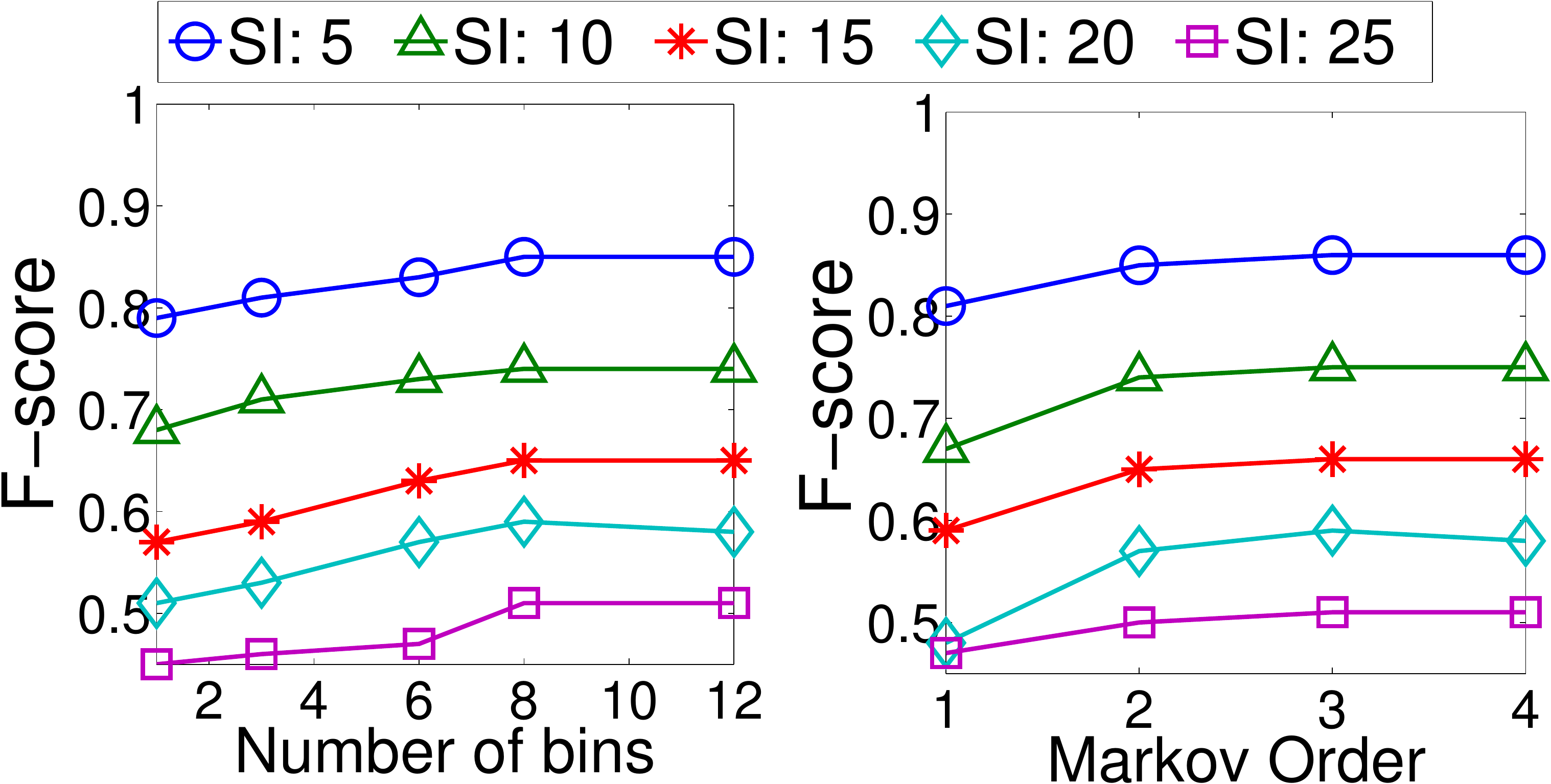}}
		\subfigure[]{
	\label{fig:markovVsTime}
		\includegraphics[width=1.10in]{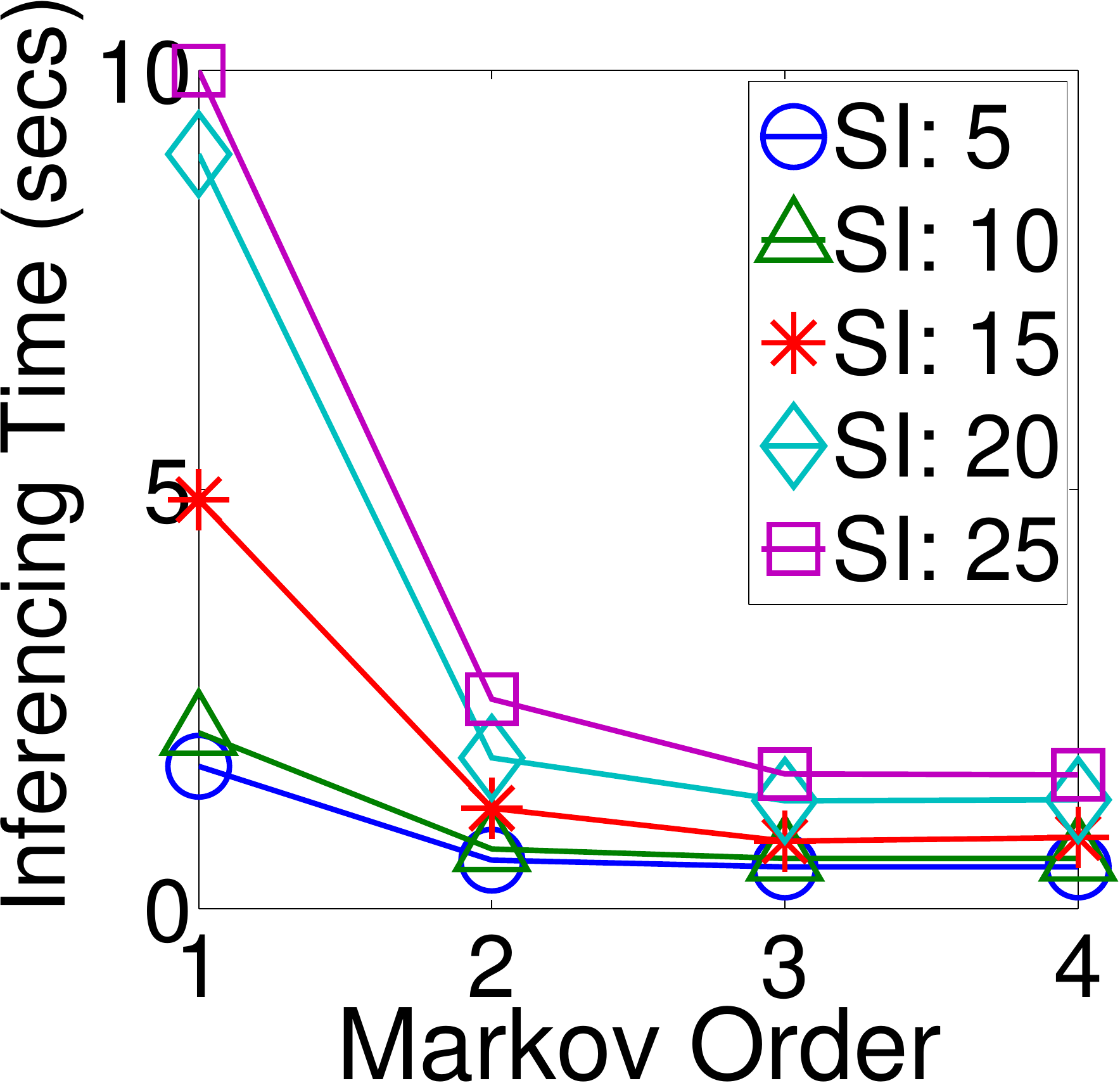}}
		\subfigure[]{
	\label{fig:markovVsSize}
		\includegraphics[width=1.07in]{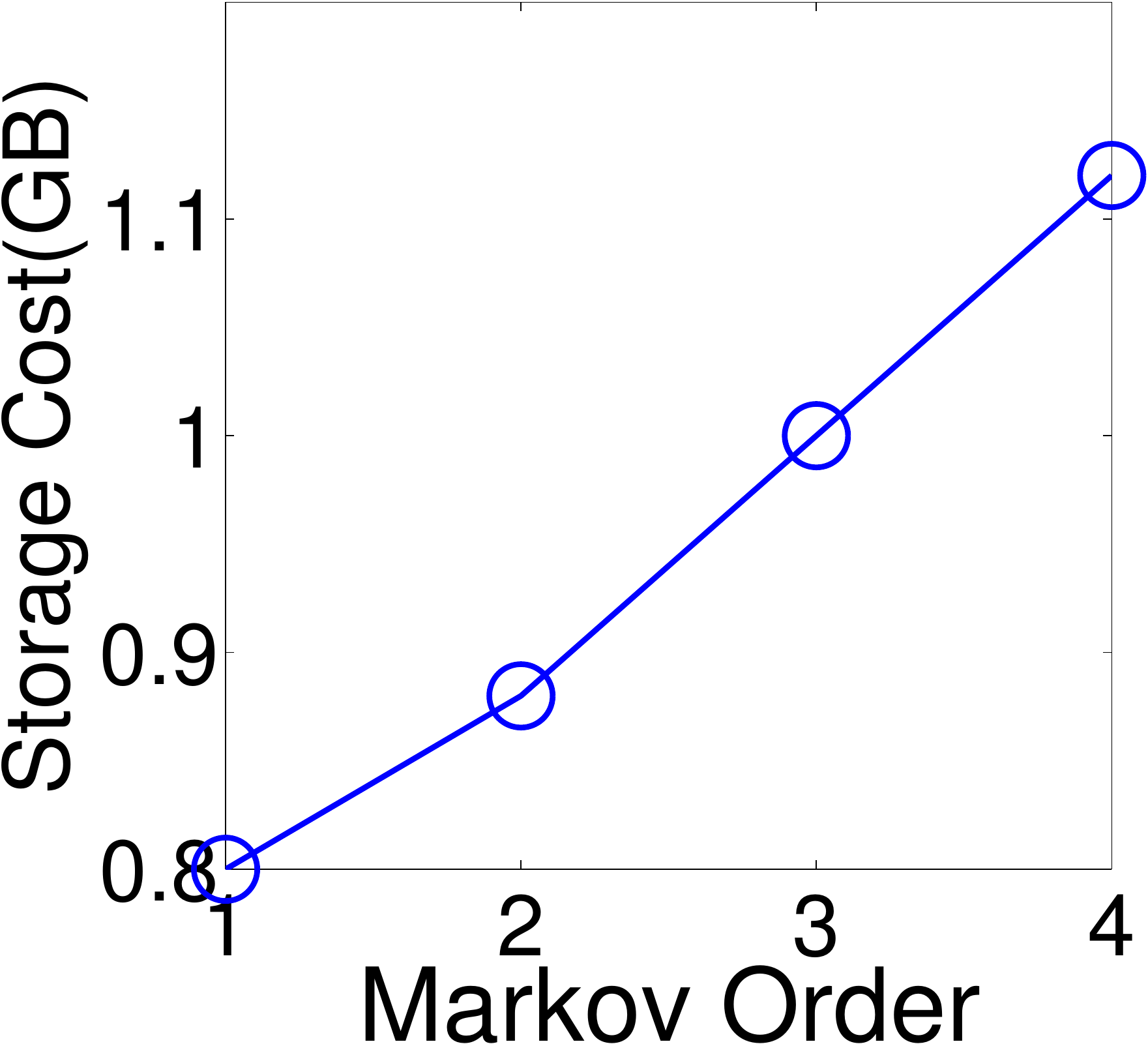}}
		\subfigure[]{
	\label{fig:spacePrediction}
		\includegraphics[width=1.07in]{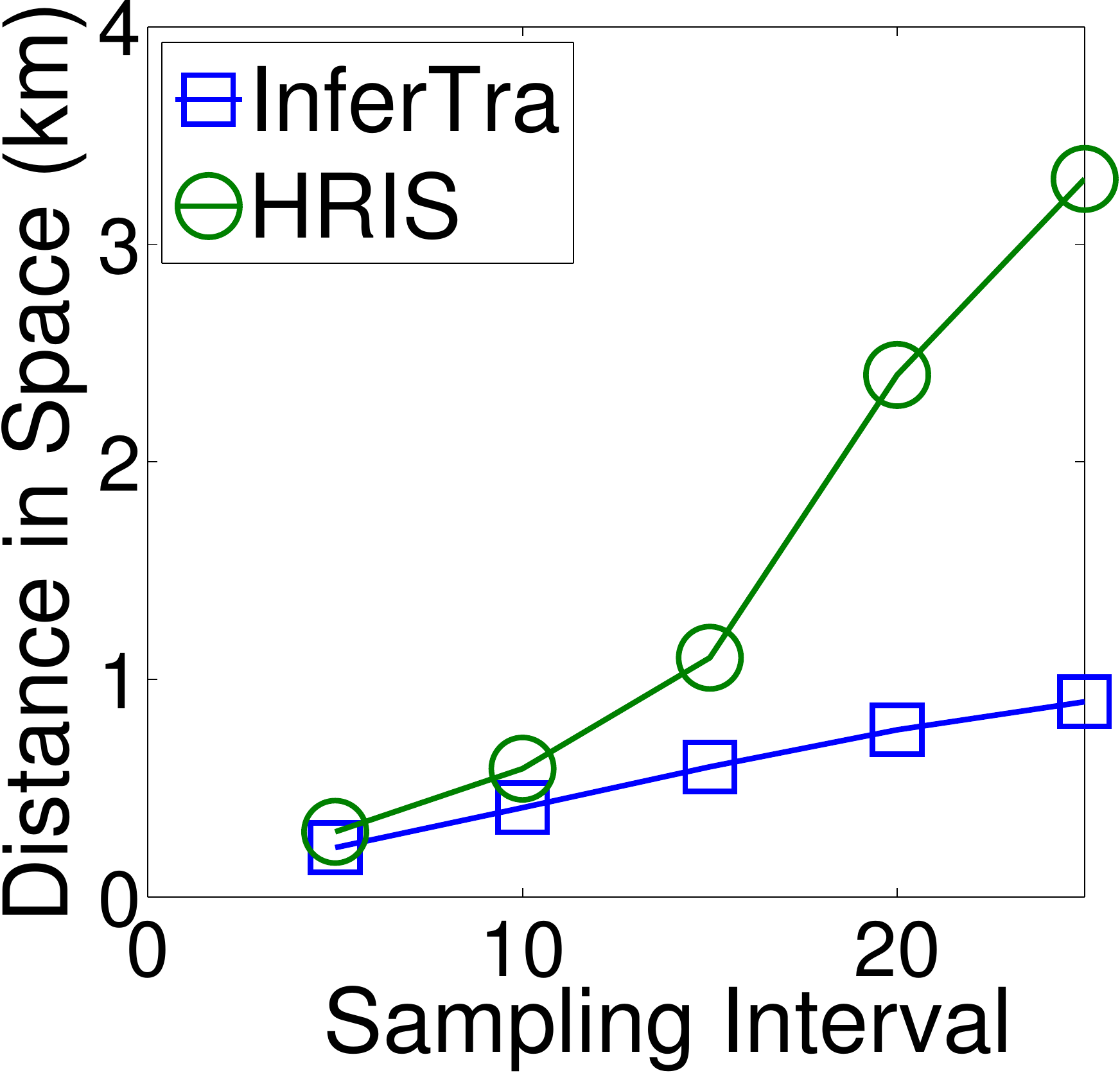}}
		\subfigure[]{
	\label{fig:timePrediction}
		\includegraphics[width=1.07in]{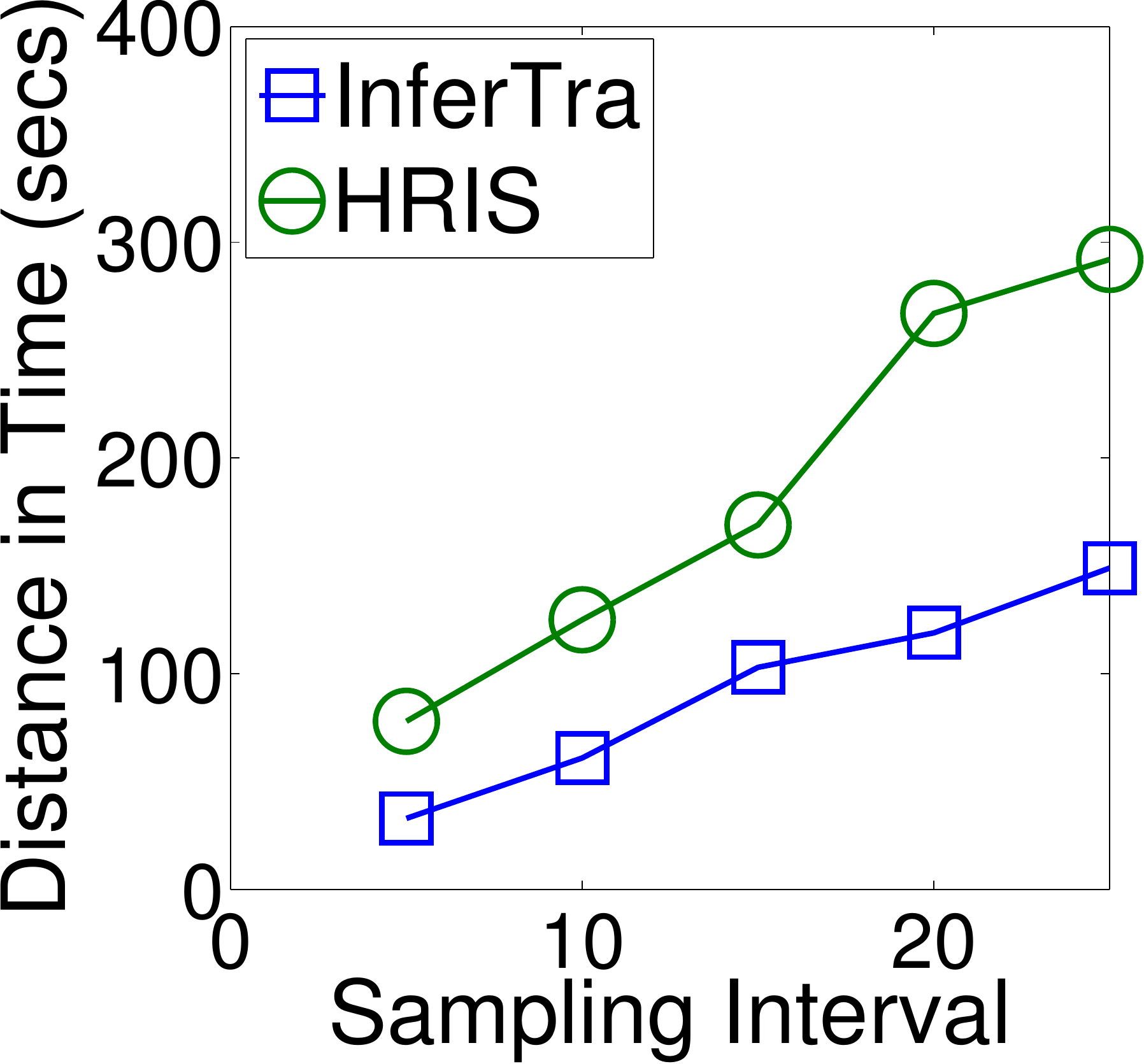}}
		\subfigure[]{
	\label{fig:distanceVsAccuracy}
		\includegraphics[width=1.07in]{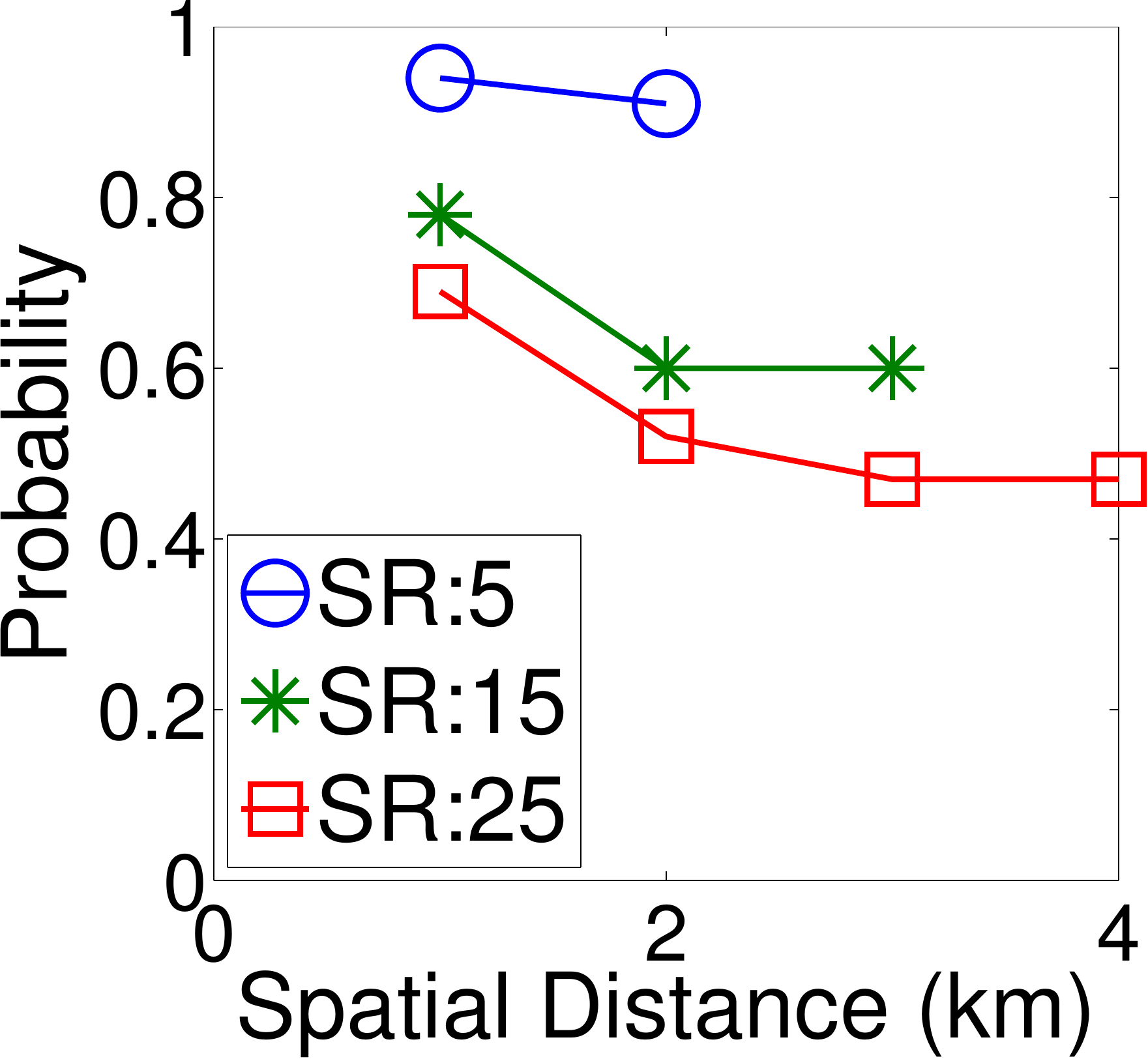}}
		\subfigure[]{
	\label{fig:timeDistanceVsAccuracy}
		\includegraphics[width=1.10in]{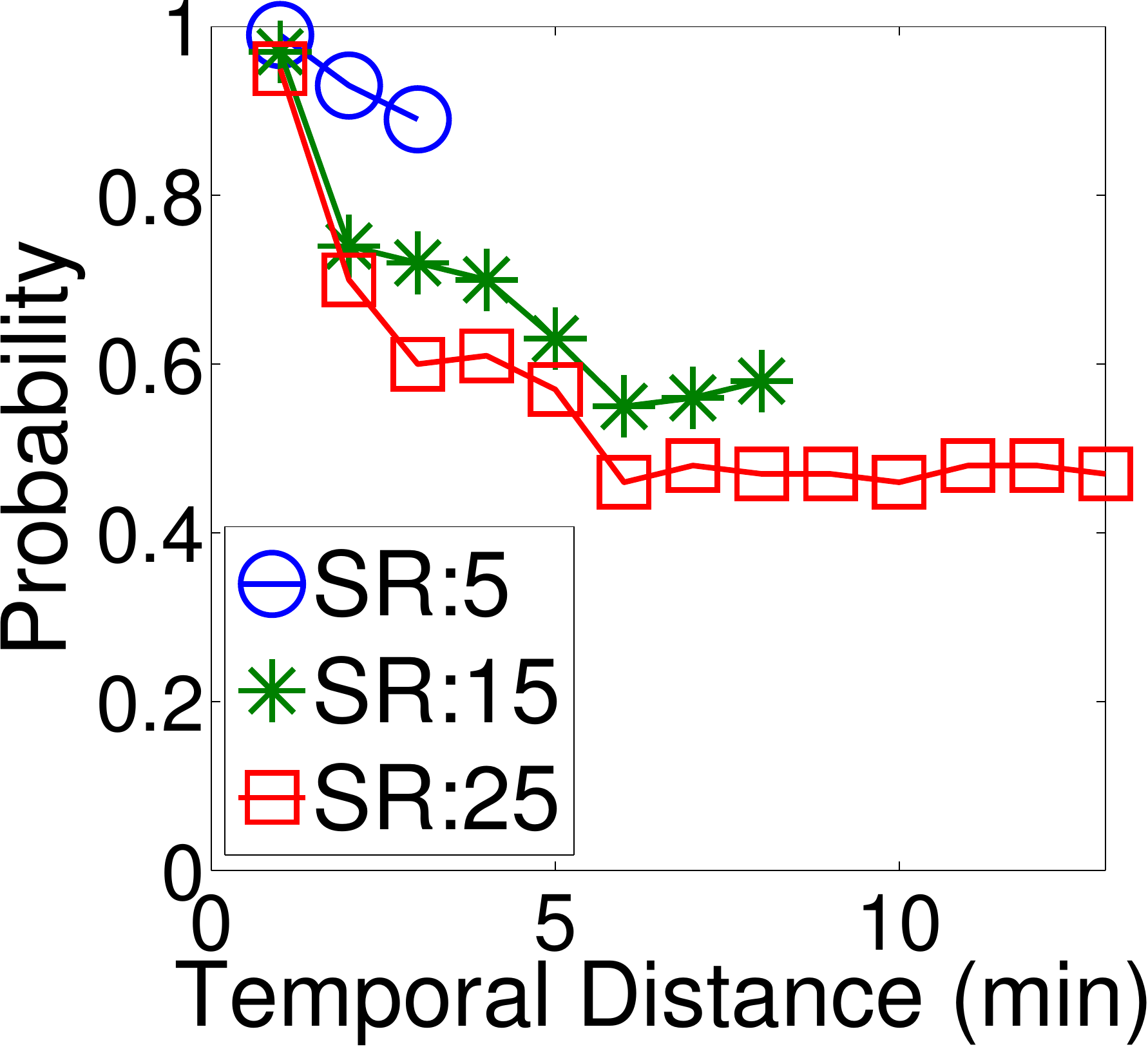}}
\caption{Growth rate of accuracy and inferencing time with (a) sampling interval (SI) and (b-c) training dataset size. (d) Growth rate of accuracy with  number of bins in the affinity vector, and order of the Markov model. Growth rate of (e) inferencing time, and (f) storage with the order of the Markov model. (g) Spatial and (h) temporal distances between actual times and actual locations, respectively, against sampling interval. Node probability against distance in (i) space and (j) time from the nearest observation.}
\end{figure*}
\subsection{Performance of InferTra}
\label{sec:infertraperformance}
Among the various queries outlined in Sec.~\ref{sec:introduction}, we first benchmark the performance on trajectory inference.
\subsubsection{What is the actual trajectory?}
First, we study the impact of sampling interval (SI) on trajectory inference. The first plot in Fig.~\ref{fig:samplingRateVsAccuracy} demonstrates the results as the SI is varied from $5$ minutes per point to $25$ minutes. As expected, the F-score decreases with increase in the SI. Across all SIs, InferTra outperforms HRIS.  STP does not display a good performance since drivers do not have a global knowledge of the shortest routes. Furthermore, a separate work has shown that people prefer more pleasant routes than the quickest~\cite{happyroute}. In addition to the uncertain trajectory predicted by InferTra, we also evaluate its maximum likelihood trajectory (MLT) computed as outlined in Sec.~\ref{sec:properties}. The accuracy of the MLT is comparable to the uncertain trajectory till an SI of $10$. Beyond $10$, there is a sharp deterioration and it resembles the accuracy of HRIS. This result highlights why it is important to go beyond maximum likelihood estimations and capture the entire uncertainty surrounding partial observations. Our more holistic approach to inference not only ensures a higher accuracy, but also enables slower deterioration rate with SI. Consequently, even at an SI of $25$ minutes, InferTra achieves an F-score of $0.51$, which is $50$\% higher than the F-score of $0.32$ by HRIS. 

Next, we benchmark the inferencing time of InferTra. Plot two in Fig.~\ref{fig:samplingRateVsAccuracy} demonstrates the results. 
%Naturally, the inference time increases with SI. 
At a higher SI, there is more uncertainty between two intermediate points, and consequently, a higher inference time is required. While SP is marginally faster than InferTra, InferTra is up to 20 times faster than HRIS. This amazing speed-up is achieved since InferTra is only reliant on local state transitions. In contrast, HRIS performs a search across the entire dataset to identify trajectories that overlap with the given observations. Based on the results, a relevant subgraph is extracted on which the inference is made. Due to the online searching of the trajectory database, a high inferencing time is incurred in HRIS.

Continuing on the topic of inference time, we next study how it varies with the training dataset size. As shown in Fig.~\ref{fig:trainingSizeVsTime}, InferTra has the attractive property of a decreasing inference time with increase in training data. As more training data is available, the conditional probabilities get more accurate and consequently, converges quicker to the joint distribution. On the other hand, the inferencing time of HRIS grows with the training data size since the online search on the trajectory database imparts a larger cost. In addition to the inferencing time, we also study the impact of training data on the accuracy. As expected, the performances of both techniques improve with additional training data.

InferTra not only learns spatial signals, but is also sensitive to any temporal periodicity in the mobility patterns. We thus analyze if identifying such temporal periodicities improves the inference performance. Towards that goal, instead of automatically learning the optimal number of bins in the affinity vector at each edge, we manually set the number of bins across all edges. Lower the number of bins, the more is the reliance on using only the spatial signals. In the extreme case where only $1$ bin is used, no temporal information is incorporated in the NMM. Fig.~\ref{fig:markovVsAccuracy} demonstrates the results. With increase in the number of bins, there is up to 10\% increase in the F-score. Across all SIs, the improvement saturates at $8$ bins. This result clearly highlights the importance of capturing temporal patterns in addition to the spatial signals. Prior to InferTra, this dimension of the inference problem has not been explored.

\textit{Do vehicles take locally optimal decisions in route selections?} The order of the Markov model in the NMM controls how well this preference is learned. We next study this issue. First, we evaluate the inference accuracy as a longer history is retained to determine the transition probabilities in the NMM. As it can be seen in Fig.~\ref{fig:markovVsAccuracy}, the F-Score saturates at an order of $3$ across all SIs. This result, combined with the weak performance of STP, show that vehicles typically make locally optimum decisions and thus, looking too far back in the history is not useful. 
A trend similar to the accuracy is also visible in the inferencing time analysis in Fig.~\ref{fig:markovVsTime}. At an order of $1$, where only the current state is examined probabilities, the mobility patterns are not captured as accurately. Due to the resultant ambiguity, the convergence to the joint distribution is slower. 
%At an order of $2$ onwards, the patterns are more accurately captured, which improves the inference times. 
Although, a higher Markov order incurs more storage cost, as shown in Fig.~\ref{fig:markovVsSize}, the growth rate is linear. This result is expected given the fact that the average outgoing degree in the Beijing road network is close to $1$. 
Overall, to summarize the attractive features of InferTra:
\begin{itemize*}
  \item By not relying on shortest paths, which is at the core of HRIS, InferTra is up to $50$\% more accurate and $20$ times faster.
  \item Both the efficiency and the accuracy of InferTra improve with increase in the size of training data.\
  \item InferTra is the first technique to capture the temporal signals embedded in historical trajectories.
  \item InferTra has only $2$ parameters, of which only $1$ has any noticeable impact on the performance.
  \end{itemize*}
\subsubsection{Where should I search at time $t$?}
Given a set of observations, what was the most likely location of a vehicle at time $t$? Or, if a vehicle was at node $n$, when did the vehicle reach $n$? Such queries routinely find applications in retrospective analysis of crime investigations, and existing techniques cannot answer them adequately.  We analyze performance of InferTra in these scenarios. Recall, Alg.~\ref{alg:sampletraj} also estimates the time at any node $n$ in the uncertain trajectory (line 14).

First, we study the accuracy of the predicted location at an input time $t$. To setup the experiment, the input time $t$ is set to the timestamp of a randomly picked node that is part of the ground truth trajectory, but not included in the observation set. We quantify the prediction accuracy, by computing the spatial distance between the actual and the predicted locations at time $t$. In InferTra, there can be multiple routes to the destination and therefore, a distribution of nodes is produced as possible locations at time $t$. We compute the overall spatial distance $sd$, by taking their weighted sum. More formally,
\begin{equation}
	sd=\sum_{\forall d_n\in\mathbb{SD}}p(d_n)\times d_n
      \end{equation}
where $\mathbb{SD}$ is the set of distances from the actual location corresponding to each predicted node $n$ at time $t$, and $p(d_n)$ is the associated probability.% of the $i_{th}$ node in $\mathbb{SD}$. 

 In its original form, HRIS cannot answer this query. First, HRIS cannot predict time.  In addition, to predict location at an input time, one needs to analyze node-level likelihoods instead of whole trajectory likelihoods. Nonetheless, for benchmarking purposes, we extract an answer from HRIS based on the assumption that the most likely node at time $t$ is part of the most likely trajectory. We estimate the time at each node based on the average speeds in its constituent edges. Fig.~\ref{fig:spacePrediction} shows the results as the SI is varied. While InferTra estimates the location within a radius of 1 KM even at an SI of $25$ minutes, the error range in HRIS is as high as $3.5$ KM. This result stems from the fact that the most likely node may not necessarily be part of the most likely trajectory. While an uncertain trajectory captures both node and trajectory level likelihoods, HRIS relies on a close correspondence between the most likely trajectory and most likely node. Consequently, the error range is higher. 

 Fig.~\ref{fig:timePrediction} performs the dual of the previous study. Instead of predicting the location at a given time, we predict the time at a given node $n$ from the ground truth. If the input node $n$ is not part of the uncertain trajectory or the HRIS-inferred trajectory, we output the time at the node that is spatially closest to $n$. Similar to the previous query, HRIS is not built for answering this query. This inability to make node-level predictions coupled with the tight integration of both spatial and temporal patterns in the inference procedure of InferTra, result in a significant performance disparity. As visible in Fig.~\ref{fig:timePrediction}, InferTra is 2 times more accurate than HRIS.

 Finally, we look at InferTra's performance in predicting node likelihoods based on its distance from the closest observation. To assess the performance, we randomly pick nodes from ground truth trajectories and compute its visit likelihood in the InferTra prediction. Generally, closer a node is to an observation, higher is its probability; as the distance from the observation grows, the more difficult the prediction task becomes. Figs.~\ref{fig:distanceVsAccuracy} and \ref{fig:timeDistanceVsAccuracy} demonstrate the results against spatial and temporal distances respectively. The trends are similar and in line with the general intuition. It is interesting to note that the SI is an important factor. To give an example, when the destination is 10 minutes away, the number of possible routes between the source and the destination is much larger. Consequently a higher number of candidate nodes are present that are 1 KM away. On the other hand, if the destination is only 2 minutes away, only a few routes exist that can connect the source to destination within the observed time. As a result, the candidate space is smaller, and hence, less is the uncertainty.
%Fig.~\ref{fig:timePrediction} demonstrates the distance between the predicted time and the actual time on a node $n$ that is part of the ground truth trajectory, but not the observation set. The predicted time is computed based on Alg.~\ref{alg:sampleTrajectory} (line 14). Note that there can be multiple routes to $n$ with each route having their own predicted time. Thus, the overall distance in time, $td$, is computed as the weighted summation of the temporal distances in each route. More formally,
\section{Conclusion}
  \label{sec:conclusion}
  In this paper, we studied the problem of trajectory inference from partial observations. We developed a technique called \emph{InferTra} that summarizes all of the possibilities through an ``uncertain'' trajectory. By taking the shape of an edge-weighted graph, an uncertain trajectory captures a richer representation of the uncertainty surrounding partial observations than maximum likelihood estimations. InferTra is built on the foundation of Gibbs sampling and is powered by a \emph{Network Mobility Model (NMM)}, which not only utilizes the spatial patterns embedded in historical data, but also unearths how these patterns vary with time. Extensive experiments on real network-constrained trajectories showed InferTra to be up to $50$\% more accurate and $20$ times faster than the state-of-the-art inferencing technique. In addition, an uncertain trajectory can handle a wider range of important queries. 
  %To further unleash the potential of uncertain trajectories, we designed a framework called \emph{gDTW}, which allows trajectory matching in the uncertain world. Without compromising on the running time, gDTW is up to $2$ times more accurate than applying DTW on maximum likelihood trajectories.

%Consequently, InferTra predicts an edge-weighted graph, which is a richer representation of the uncertainty surrounding the partial observations than the most likely trajectory. More importantly, the uncertain trajectory can be used to answer all of the queries highlighted above.
  {\scriptsize
  \bibliographystyle{IEEEtran}
  \bibliography{uTraj}}
\end{document}